\documentclass[12pt]{spieman}  
\usepackage{amsmath,amsfonts,amssymb}
\usepackage{graphicx}
\usepackage{setspace}
\usepackage{tocloft}
\usepackage{subcaption}
\usepackage{caption}
\usepackage{deluxetable}

\newcommand{\HWO}{\emph{Habitable Worlds Observatory}}

\newcommand{\mearth}{\ensuremath{M_{\oplus}}}

\newcommand{\Rearth}{\ensuremath{R_{\oplus}}}

\newcommand{\JPL}{Jet Propulsion Laboratory, California Institute of Technology, 4800 Oak Grove Dr., Pasadena, CA 91109, USA}
\newcommand{\umich}{University of Michigan, Department of Astronomy, 1085 South University, Ann Arbor MI, 48109, USA}
\newcommand{\UWMadison}{Department of Astronomy, University of Wisconsin-Madison, 475 N Charter St, Madison, WI 53706, USA}
\newcommand{\AMES}{NASA Ames Research Center, Moffett Field, CA 94035, USA}
\newcommand{\UA}{Department of Astronomy and Steward Observatory, University of Arizona, Tucson, AZ 85721, USA}
\newcommand{\NMSU}{Department of Astronomy, New Mexico State University, P.O. Box 30001, MSC 4500, Las Cruces, NM 88003, USA}

\title{Searching for Habitable Exoplanets with Relative Astrometry (SHERA). I. The Case for Searching for Planets in Binary Star Systems}

\author[a,*]{Jessie L. Christiansen}
\author[b]{Eric E. Mamajek}
\author[b]{Gautam Vasisht}
\author[a]{Catherine A. Clark}
\author[c]{William Roberson}
\author[c]{Eric L. Nielsen}
\author[d]{Kaitlin M. Kratter}
\author[e]{Juliette Becker}
\author[f]{Eduardo Bendek}
\author[f]{Ruslan Belikov}
\author[b]{Alex Davis}
\author[g]{Louis Desdoigts}
\author[e]{Alyssa Jankowski}
\author[h]{Michael R. Meyer}
\author[i]{Benjamin J. S. Pope}
\author[b]{Armen Tokadjian}
\author[j]{Peter Tuthill}
\affil[a]{NASA Exoplanet Science Institute, IPAC, MS 100-22, Caltech, 1200 E. California Blvd, Pasadena, CA 91125}
\affil[b]{\JPL}
\affil[c]{\NMSU}
\affil[d]{\UA}
\affil[e]{\UWMadison}
\affil[f]{\AMES}
\affil[g]{Leiden Observatory, Niels Bohrweg 2, Leiden 2300RA, The Netherlands}
\affil[h]{\umich}
\affil[i]{School of Mathematical \& Physical Sciences, Macquarie University, 12 Wally\'s Walk, Macquarie Park, NSW 2113, Australia}
\affil[j]{Sydney Institute for Astronomy, School of Physics, University of Sydney, NSW 2006, Australia}

\cftpagenumbersoff{figure}
\cftpagenumbersoff{table} 
\begin{document} 
\maketitle

\begin{abstract}
Discovering Earth-like planets orbiting Sun-like stars was identified as a priority science goal of the Astronomy 2020 Decadal Survey. It is confounded by many factors, one of which is the high multiplicity of Sun-like stars in the local neighborhood---half of nearby Sun-like stars are in binary or higher-order stellar systems, which are less amenable to the detection of small planets with almost all of the currently productive exoplanet detection techniques. Here we describe the SHERA (Searching for Habitable Exoplanets with Relative Astrometry) NASA Small Explorer mission concept. SHERA utilizes diffractive-pupil technology on a small, simple optical space telescope to achieve microarcsecond precision relative astrometry on 14 Sun-like stars in seven nearby multi-star systems, combining the pupil and stellar binarity to provide a precise reference in the image plane. With this precision, SHERA would enable: (i) a search for rocky planets in the habitable zones of the closest Sun-like stars; (ii) an investigation of the impact of binary star formation on small, widely separated planets; and (iii) the performance of crucial precursor observations on a number of high-priority targets of NASA's future missions to characterize Earth-like planets, such as the \HWO. When combined with radial velocity measurements, SHERA relative astrometry will also enable exploration of the three-dimensional orbital structure of planets in binary systems.
\end{abstract}


{\noindent \footnotesize\textbf{*} Jessie L. Christiansen,  \linkable{jessiec@caltech.edu} }

\begin{spacing}{2}   

\section{Introduction}
\label{sect:intro}  
The discovery and characterization of planets orbiting other stars has been a remarkable success. In 2026, over 6000 confirmed or statistically validated exoplanets are known, as well as thousands of additional planet candidates \cite{Christiansen2025}. This dramatic increase has been driven by advances across multiple astronomical measurement techniques, including precision radial velocity measurements, long term photometric monitoring, as well as improvements in data detrending techniques. Much of this progress has been driven by the goal of achieving the precision required to detect and characterize Earth-like planets (for the purposes of this paper, ``Earth-like planets'' are considered to be rocky planets in the habitable zones [HZ] of Sun-like stars). However, these planets remain largely beyond the limits of our detection techniques. In this paper, we present a concept for a small (Explorer-class) astrometry mission, SHERA (Searching for Habitable Exoplanets with Relative Astrometry) to search for rocky, habitable-zone planets around Sun-like stars in nearby binary and higher-order multi-star systems, as well as to perform important precursor science to help inform the design and mission planning for the NASA \HWO\ mission.

In Section \ref{sec:stateoftheart}, we describe the current state-of-the-art precision achieved in detecting Earth-like planets, as well as near-future plans for improving on the current limits. Section \ref{sec:relastro} describes the opportunity afforded by relative astrometry, with respect to nearby Sun-like stars in binaries and higher-order multi-star systems. In Section \ref{sec:binaries} we discuss the prospects for planet formation and evolution in binary systems, including the current open questions. Section \ref{sec:SHERA} outlines the SHERA mission concept, including the science objectives and potential ancillary science. Section \ref{sec:hwo} outlines the precursor science that a mission like SHERA would contribute to planning and target selection for the future NASA \HWO\ flagship mission. 

\section{State-of-the-Art for Earth-Like Planet Detection}
\label{sec:stateoftheart}

There are several methods in use today for the detection of Earth-like exoplanets, including transit, radial velocity, microlensing, direct imaging, and absolute astrometry. Here we discuss the state-of-the-art for each.

\subsection{Transit}

When transiting, Earth-sized planets block $\sim$100 ppm of the light from a Sun-like star. The NASA {\it Kepler} mission \cite{Borucki2010,Koch2010} was a single 1-m optical telescope which observed a single field-of-view (FOV) continuously from 2009--2013, with the primary goal of measuring the occurrence rate of Earth-sized planets in the HZ of Sun-like stars ($\eta_{\oplus}$). To achieve this goal, it was engineered to obtain time series photometry with a noise level over 6.5 hours of $\sim$20 ppm for a $V\simeq12$ mag star, however stars and instrumental systematics combined to increase the typical measured noise level for these stars to 32 ppm, significantly higher than expected \cite{Gilliland2011}. Ultimately, there have been no confirmed detections of Earth-like planets orbiting Sun-like stars with {\it Kepler}, however it remains the state-of-the-art for photometric precision over year-long observing baselines. The NASA {\it TESS} mission \cite{Ricker2015}, operating since 2018, consists of 4 10-cm optical telescopes, observing the sky in 27-day sectors. Although there are two continuous viewing zones that have received up to 351 days of continuous observations, {\it TESS} has neither the photometric precision nor the observing baseline to detect Earth-like planets (nor was that the goal of the mission).

Upcoming transit surveys are planning to detect Earth-like planets, in either small numbers or around quite faint stars. The European Space Agency's {\it Plato}\footnote{PLAnetary Transits and Oscillations. \url{https://www.esa.int/Science_Exploration/Space_Science/Plato}} mission \cite{Plato2025}, launching in 2027, consists of 24 partially overlapping 12-cm optical telescopes, and is expected to achieve $\sim$34 ppm for 11th magnitude stars in its deepest overlap. Over four years it will perform 1--2 long stares of 2--3 years each, and is expected to find $\sim$1--3 small ($<1.5 \Rearth$) HZ planets around its brightest ($m_{\rm V}\sim8$) targets \cite{Heller2022}, and 8--10 around fainter targets down to $m_{\rm V}=11$. All of these potential detections will be too distant ($>20$ pc) for direct imaging follow-up observations with any of the planned next generation imaging facilities, and with atmospheric scale heights that are well below the detection limits for current or future generations of instruments that could obtain transmission spectroscopy during transit. The Chinese Space Agency's {\it Earth 2.0} mission \cite{Ge2024}, launching in 2028, similarly consists of six partially overlapping 28-cm optical telescopes, and it will perform a four-year survey centered on the {\it Kepler} field. It is predicted to find 10--20 Earth-like planets; although the magnitude range of the potential host stars is not published, they are likely to be fainter on average than {\it Plato}, given the respective telescope sizes (larger for {\it Earth 2.0}) and fields of view (smaller for {\it Earth 2.0}). As a result, they are also expected to be too distant for direct imaging follow-up or transmission spectroscopy observations. For both missions, the science objective is to constrain the frequency of Earth-like planets ($\eta_{\oplus}$), rather than identify individual planets for detailed characterization.

\subsection{Radial Velocity}

The reflex motion of a Sun-like star induced by an Earth-mass planet in the HZ is $\sim$9 cm/s. As summarized in a recent review \cite{Burt2025}, the current generation of precise radial velocity (RV) instruments are now achieving $\sim$50 cm/s precision over the years-long observing baselines needed to detect Earth-like planets; as yet, none are capable of overcoming the noise floor created by stellar jitter to enable the detection of Earth-like planets. Tab. 1 of \cite{Burt2025} lists the current and upcoming RV instruments capable of achieving single-measurement precisions of $<$1 m/s at visible wavelengths. The push towards improved RV precision will likely require improvements in instrumentation, survey design, and data analysis techniques \cite{Crass2021}.

\subsection{Microlensing}

Currently, ground-based microlensing surveys have not achieved the photometric precision and cadence on sufficient targets to detect Earth-sized planets in the HZ of Sun-like stars. The Korea Microlensing Telescope Network (KMTNet)\cite{KMTNet2016} has detected two small planets (2.7--4.7$M_{\oplus}$) orbiting K stars \cite{Han2022,Zang2023}, but both lie well beyond the habitable zone. The Optical Gravitational Lensing Experiment (OGLE)\cite{OGLE2003} recently announced a low-mass ($\sim$1.3$M_{\oplus}$) planet orbiting a late K star \cite{Zang2025}, but also at a quite wide ($\sim$10 au) separation. In the future, the NASA {\it Nancy Grace Roman Space Telescope} mission, launching in 2026, will perform a microlensing survey of the Galactic Bulge \cite{Penny2019}, using a 2.4-m near-infrared telescope and its Wide Field Imaging (WFI) instrument to observe six 72-day seasons over five years. Even moreso than {\it Plato} and {\it Earth 2.0}, {\it Roman}'s science objective with this survey is planet demographics rather than individual characterization. {\it Roman} will be sensitive to Earth-like planets, with a predicted yield of 140--160 planets \cite{ROTAC2025}. However, these planets will orbit substantially fainter targets ($17<m_{\rm I}<22$), and after the conclusion of the microlensing event, the only planned and plausible further characterization will be a more precise planet mass measurement as the lens and source separate on the sky with time. The {\it Earth 2.0} mission described above also plans to have one additional 35-cm telescope observing the Galactic Bulge simultaneously with the transit survey of the {\it Kepler} field, and hopes to find up to 10 additional cold Earth-mass planets \cite{Ge2024} around these very faint stars.

\subsection{Direct Imaging}

The light contrast ratio between a Sun-like star and Earth-like planet is $\sim$2$\times$$10^{-10}$ \cite{Currie2023}, for a separation on the sky of $0.1''$ at a distance of 10 pc. The current state-of-the-art direct imaging instruments utilize coronagraphs behind extreme adaptive optics on the largest (8--10 m) ground-based telescopes to achieve $10^{-6}$--$10^{-7}$ at separations of $0.5''$ (GPIES)\cite{Nielsen2019}, (SHINE)\cite{Langlois2021}. This limits the discovery regime to $> 2 M_{Jup}$, typically at $\gtrsim$10 au separation, for young, self-luminous planets. The next generation of extremely large ground-based telescopes will include instruments that hope to push this contrast to $10^{-8}$, including the Planetary Systems Imager on the Thirty Meter Telescope (TMT/PSI)\cite{JensenClem2021} and the Planetary Camera and Spectrograph on the European Extremely Large Telescope (E-ELT/PCS)\cite{Kasper2021}. This will enable detections of warm Neptunes, or Earth-sized planets in the habitable zones of M dwarfs, and potentially a small ($<$10) number of Earth-like planets in thermal emission. However, achieving contrasts of $10^{-9}$--$10^{-10}$ likely requires a large telescope in space to overcome the noise floor in the speckle suppression caused by Earth's atmosphere. NASA's 2.4-m Nancy Grace Roman Space Telescope, planned for launch in 2026, will fly with a coronagraph instrument that plans to demonstrate a contrast of $\ge1\times10^{-7}$ for a star with $V_{AB}$ magnitude $\le$5 \cite{Pober2025}, with a further goal of pushing this to $\ge1\times10^{-9}$ and sensitivity cool Jupiter-like planets. NASA's \HWO \cite{Feinberg2026}, planned for launch in the 2040's, will survey $\sim$100 Sun-like stars with a high-contrast ($\ge1\times10^{-10})$ coronagraph on a large (6--8-m) ultraviolet/optical/infrared telescope to detect and characterize $\sim$25 Earth-like planets. ESA's Large Interferometer For Earths (LIFE) mission \cite{Quanz2022}, also planned for the 2040's, is expecting a similar yield of $\sim$25--40 Earth-like planets with four 2-m telescopes providing mid-infrared interferometry. 

\subsection{Absolute Astrometry}
\label{sec:abastro}

\begin{figure}
\begin{center}
\begin{tabular}{c}
\includegraphics[height=8cm]{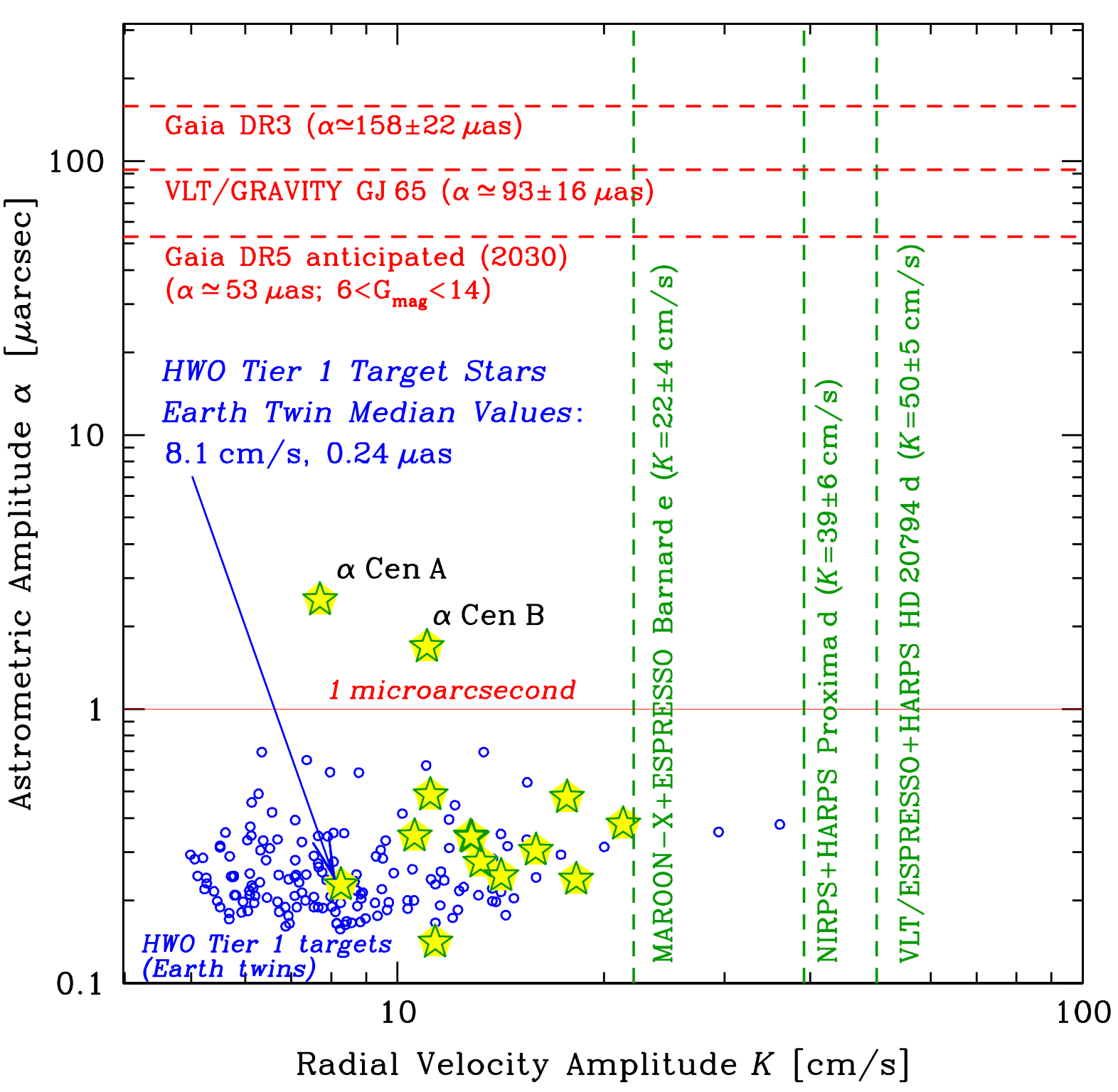} \includegraphics[height=8cm]{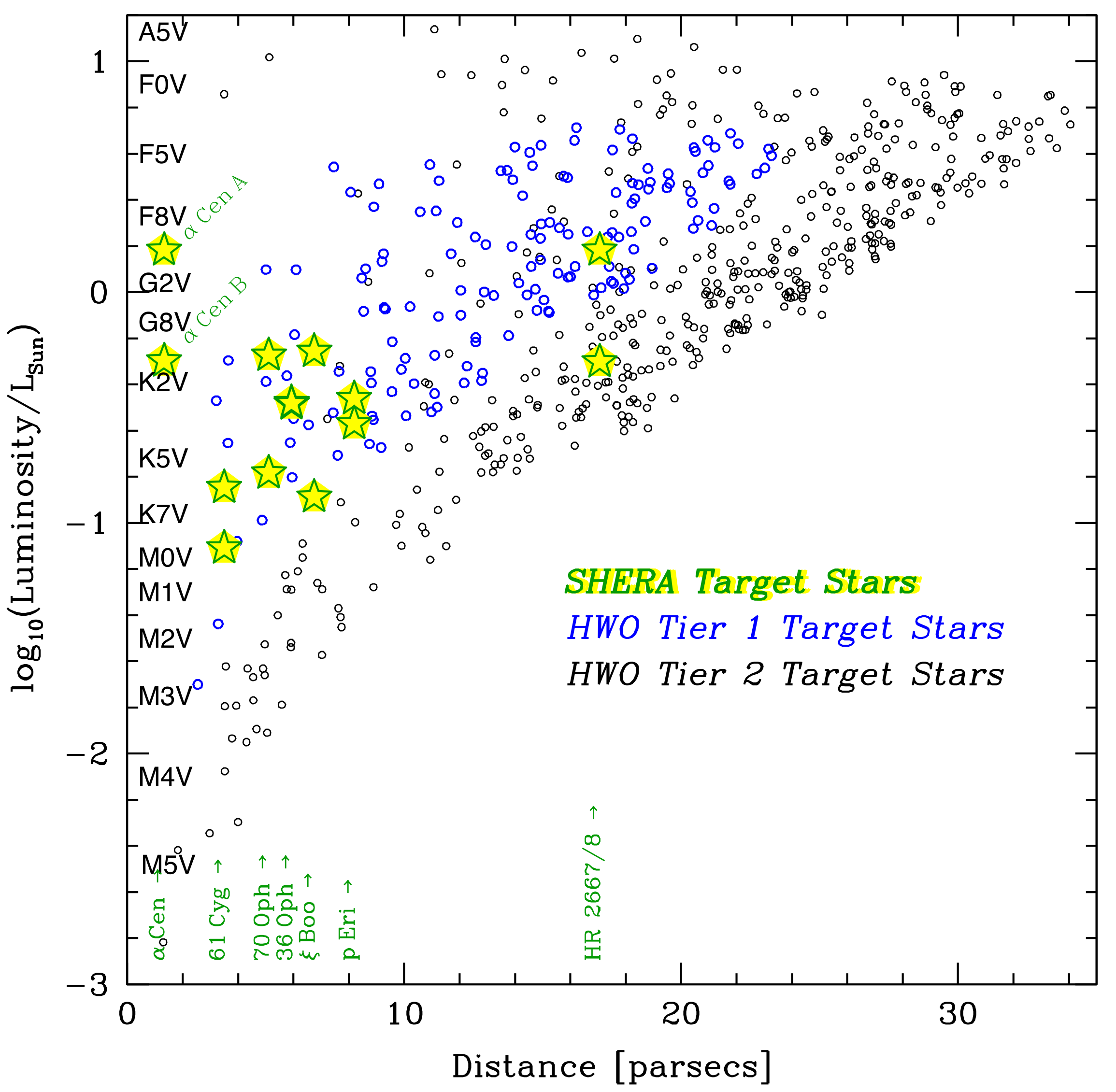}  
\\
\end{tabular}
\end{center}
\caption 
{{\it Left:} Plot of state-of-the-art detection thresholds for extreme precision radial velocity (vertical green lines) and astrometry (horizontal red lines), along with predicted amplitudes for hypothetical Earth twins orbiting at the Earth-Equivalent Instellation Distances (EEIDs) around HWO Tier 1 target stars \cite{Mamajek2024} (blue points) and the SHERA targets (green-bordered yellow stars, see Table 1). The benchmark level of 1\,$\mu$as is shown as a solid red line. 
{\it Right:} Distance versus bolometric luminosity for SHERA targets (green-bordered yellow stars), HWO Tier 1 \cite{Mamajek2024} (blue points) and Tier 2 \cite{Tuchow2025} (grey points) target stars.
\label{fig:astrometryRV}} 
\end{figure}

The maximum astrometric amplitude induced by gravitational perturbations upon a star by an orbiting companion in a circular orbit is

\begin{equation}
\alpha\, = \frac{M_{p}}{M_{*} + M_{p}}\left(\frac{a}{d}\right) \simeq\ 3.00\,\mu {\rm as}\, \left(\frac{M_{p}}{M_{\oplus}}\right) \left(\frac{M_{*} + M_{p}}{M_{\odot}}\right)^{-1} \left( \frac{a}{1\,{\rm au}}\right) \left(\frac{d}{1\,{\rm pc}}\right)^{-1}
\label{eq:astrometry}
\end{equation} 

\noindent where $M_p$ is the mass of the orbiting perturber, $M_{*}$ is the mass of the star, $a$ is the semi-major axis of the orbit in au, and $d$ is the distance to the system in parsecs (where $d$ = 1000\,$\varpi^{-1}$, where $\varpi$ is the parallax in milliarcseconds) \cite{Lammers2026}. $M_{\oplus}$ is in Earth masses and $M_{\odot}$ is in solar masses. For the nearest Sun-like stars, $\alpha$ Cen A and B at $d$ = 1.33\,pc, Earth twins orbiting these stars at the Earth-equivalent instellation distances (EEID) of 1.23 and 0.71\,au, respectively, would induce astrometric amplitudes of 2.5 and 1.7\,$\mu$as, respectively. These are by far the largest predicted amplitudes among nearby Sun-like stars, where the $d$ term in the denominator quickly dominates the numerator $a$. For an Earth twin orbiting a Sun twin at $d = 10$\,pc, a typical distance for a target stars for  imaging Earth-like planets with the proposed \HWO \cite{Mamajek2024}, the predicted astrometric amplitude would be $\sim$0.3\,$\mu$as.
The current state-of-the-art level of astrometric amplitudes that have been reported both from ground-based and space-based instruments is orders of magnitude larger than this (Figure \ref{fig:astrometryRV}).

State-of-the-art astrometric detections are currently
near the $\sim$100\,$\mu$as level, including the published performance of Gaia DR3 \cite{Holl2023}, and a $93\pm16$ $\mu$as detection for GJ 65 with VLT/GRAVITY \cite{GRAVITYcollaboration2024}.

\section{Relative Astrometry for Earth-Like Planet Detection}
\label{sec:relastro}

Traditional absolute stellar astrometry (Section \ref{sec:abastro}) is dominated by two noise sources---astrophysical (including photon noise, starspot-induced jitter \cite{Marakov2009, Sowmya2021}, stellar companions, parallax, and proper motion), and instrumental, where distortions in the optical path manifest as astrometric motion of the target in the image plane. Mitigating these effects has historically relied on balancing two competing constraints: maximizing the number of background reference stars in order to minimize the astrophysical noise, which favors a larger field of view (FoV); and minimizing the optical distortion, which to first order scales as a cubic function of the field angle \cite{Smith2008} and thus motivates a smaller FoV. For a telescope of given aperture and exposure time, an optimal FoV exists that balances these competing demands and sets the fundamental performance floor of the system \cite{Guyon2012}.

Narrow-angle relative astrometry between the components of a binary system circumvents both these limitations \cite{ShaoColavita1992, Lane2004}. As the FoV shrinks to arcsecond scales (e.g., $\sim$30$^{\prime\prime}$, rather than fractions of a degree), differential optical distortion is dramatically suppressed and approaches a pseudo-linear plate-scale ``breathing,'' which is efficiently calibrated by the diffractive pupil technology described in Section~\ref{sec:pupil}. Furthermore, the binary companion serves as the astrometric reference. The fainter the astrometric reference star/s, the higher the photon noise floor and subsequently the worse achievable astrometric precision. Two bright stars enable much improved photon-noise-limited performance, without the need for many co-added exposures, opening the door to sub-microarcsecond-class precision.

\begin{figure}
\begin{center}
\begin{tabular}{c}
\includegraphics[height=6cm]{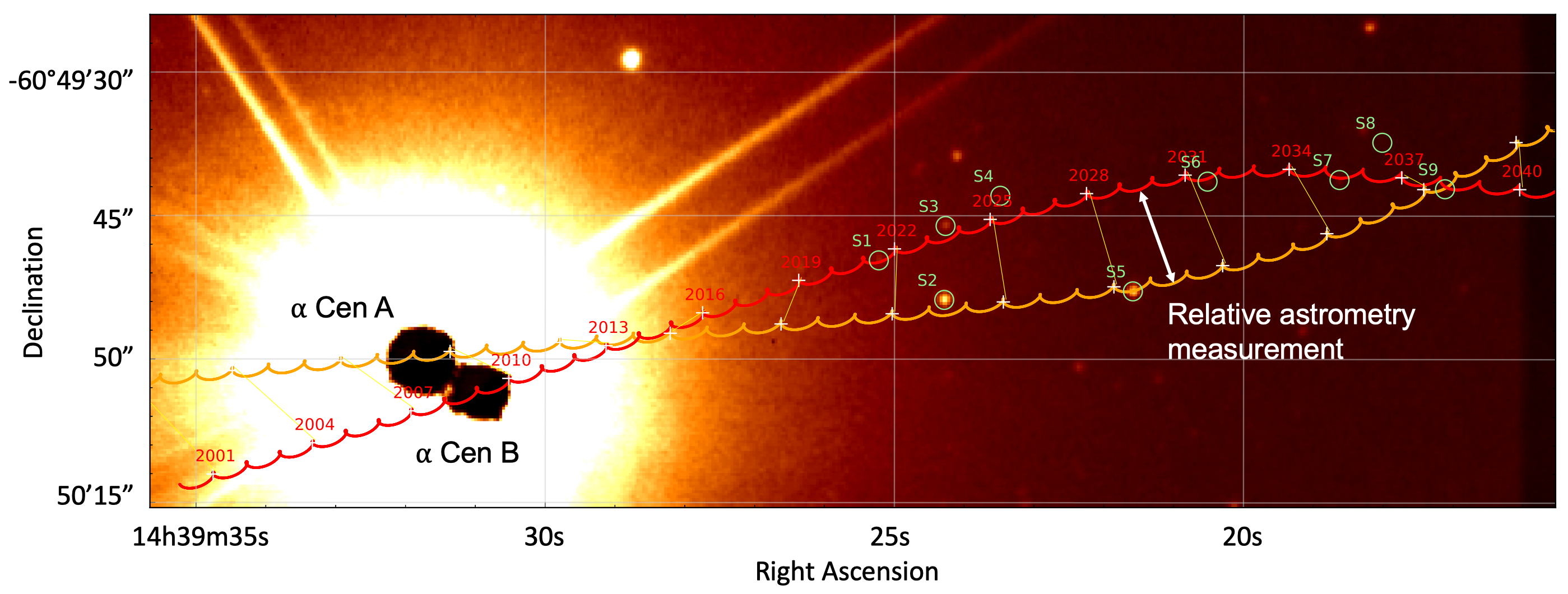}
\end{tabular}
\end{center}
\caption 
{Projected separation of $\alpha$~Centauri~A and B over time, including the parallactic wobble introduced by Earth's orbit around the Sun on one-year timescales, and the larger motion of the two stars on their $\sim$80-year orbit. The instantaneous relative astrometric vector between the two components is highlighted in white. Planetary companions would induce a periodic modulation of this vector. Courtesy: Pierre Kervella.
\label{fig:aCen_rel_astrometry}}
\end{figure}

The observable is the projected separation vector between the two stellar components as a function of time (Figure~\ref{fig:aCen_rel_astrometry}). A planetary companion orbiting either star imprints a periodic reflex modulation on this vector, revealing the presence of the planet. This technique has notable precedents: \cite{Lane2004} demonstrated differential astrometry of sub-arcsecond binaries with the Palomar Testbed Interferometer; \cite{Kok2013} extended phase-referenced narrow-angle astrometry to optical baselines with SUSI; and most recently, \cite{GRAVITYcollaboration2024} used VLTI/GRAVITY to monitor the relative astrometry of the nearest M-dwarf binary GJ~65\,AB over seven years, detecting a Neptune-mass candidate planet at $36 \pm 7\,M_\oplus$ with a mean per-epoch precision of $50$--$60\,\mu$as---directly demonstrating the power of this approach for planet detection in the solar neighborhood.

The best targets for narrow angle relative astrometry are bright systems with two or more comparable brightness stars having angular separations amenable to the telescope's field of view. Fortuitously, this includes many of the nearest Sun-like stars to our solar system (see Section \ref{sec:targets}), which are among the most valuable targets for the search for life, and therefore for precursor science for NASA's Habitable Worlds Observatory \cite{Mamajek2024, Tuchow2024}, ESA's LIFE mission \cite{Quanz2022}, upcoming generations of extremely large ground-based telescopes, and future generations of Earth-characterization missions, such as the Hybrid Observatory for Earth-like Exoplanets (HOEE)\cite{Soliman2025}. 

An important limitation of the relative astrometry technique is that the planetary reflex signal is a one-dimensional projection of the stellar wobble onto the binary separation vector. This introduces a degeneracy analogous to the $M_p\sin (i)$ ambiguity in radial velocity measurements \cite{Plavchan2015}: only the component of the planetary perturbation perpendicular to the line of sight and projected along the binary axis is recovered, leaving the true mass and orbital inclination partially degenerate without complementary observations. Additionally, the technique does not inherently identify which component of the binary hosts the planet; distinguishing the host requires either high signal-to-noise multi-epoch coverage to resolve the differential motion of each star about the system barycenter, or additional radial velocity data to recover the component of the stellar motion along the line of sight \cite{GRAVITYcollaboration2024}.

A further caveat of this approach is that, while the absolute precision of the astrometry is much improved with a bright binary reference star, the correlated instrumental sources of noise are not fully mitigated. This requires additional knowledge of the time-varying state of the instrument, which we address in Section \ref{sec:pupil}.

\section{Earth-Like Planets in Binary Star Systems}
\label{sec:binaries}

Most searches for Earth-like planets, such as {\it Kepler} \cite{Borucki2010}, have concentrated on single star systems, as additional stars typically complicate an already difficult measurement \cite{Ciardi2015,Sairam2024,Howell2024,Bergsten2026}. Thus far, the community has not obtained unambiguous detection of a true Earth analog. Binary star systems represent an exciting, mostly untapped reservoir of potential Earth analogs, given that roughly half of sun-like stars are in binaries \cite{Raghavan2010}. Indeed, the nearest star system---the $\alpha$ Centauri triple system---contains two Sun-like stars, both with dynamical stability zones large enough to harbor rocky, habitable-zone exoplanets. The nearest Earth-like planets may be in multi-star systems.

For close ($<$50 au) binary stars, the rate of short-period ($<$100 day) planets is observed to be significantly suppressed compared to single stars. \cite{Kraus2016} carried out a pioneering adaptive optics survey of {\it Kepler} Objects of Interest from 0.3--30\,\Rearth, finding strong suppression within 50\,au. Close binaries with circumprimary planets do exist (see $\nu$ Octantis, with a separation of 2.6\,au \cite{Cheng2025}, for example), however they seem to be rare. For intermediate separation binaries ($\sim$50--200\,au), the presence of a stellar companion appears to continue suppressing the occurrence of short-period ($<$100 day) planets \cite{Kraus2016,Moe2021}. However, the significance of the suppression seems to weaken with decreasing planet size \cite{Sullivan2026}, and as the binary separations increase \cite{Kraus2016,Moe2021,Sullivan2026}. For wide separation binaries ($\gtrsim$200 au), there is no observed difference in the rate of short-period planets between single and multi-star systems \cite{Bonavita2020,Moe2021}, with the implication that beyond these separations, the protoplanetary disks form and evolve in a decoupled way that effectively mimics two single star protoplanetary disks.

Besides a change in the overall occurrence rates, there is also evidence that the size distribution of planets in binaries changes. The planets in small--intermediate separation binaries are systematically {\emph smaller} than their single star counterparts \cite{Sullivan2024,Sullivan2026}, which could shift more planets into the Earth-size range compared to single stars. Notably, no existing surveys have been sensitive to this mass bin in multi-star systems. Finally, we emphasize that current surveys almost entirely exclude the period range of HZ planets. While it is not expected that the suppression rate of planets would decrease with increasing separation, such a trend cannot be ruled out from the existing data.

Taken together, the decreased size and frequency of planets in close binaries is consistent with a model in which stellar companions suppress planet formation by truncating the size or lifetime of protoplanetary disks \cite{Jang-Condell2015}. An analogous effect is seen around small stars---M dwarfs have a much lower occurrence rate of gas giants than larger stars \cite{Bonfils2013,Montet2014,Pass2023}, while at the same time having a {\it higher} rate of small ($<2\,\Rearth$) planets \cite{Dressing2015,Sabotta2021}. One possible explanation for this is that the smaller protoplanetary disk masses are insufficient to form giant planets, but that the material is therefore available to form a higher rate of smaller planets \cite{Lambrechts2019,Mulders2021}. The presence of a stellar companion could also enhance planetesimal growth rates at certain radii \cite{Silsbee2021}, also implying that the relative occurrence of small planets in binaries could be higher than around single stars. Analyzing short-period planets from {\it Kepler} and {\it K2}, \cite{HardegreeUllman2025} found that the occurrence rate of sub-Neptunes turned over below M0.5V, but that the prevalence of super-Earth planets continued to rise towards lower mass stars, indicating that even very low mass protoplanetary disks are capable of producing small planets (c.f. TRAPPIST-1\cite{Gillon2017}). In addition, if fewer large planets are formed which can then migrate inwards (as is commonly inferred to explain the population of short-period giant planets in single-star systems), then small planets could be preserved at their original locations, increasing their relative occurrence rate compared to single stars. 

Alternate models have suggested that binaries may also impact planet occurrence rates by dynamically ejecting planets that have formed \cite{Haghighipour2007}. Dynamical ejection scenarios would most dramatically affect wide separation planets. The impact on the mass function is more complex---stellar companions can directly eject planets of any mass, however if the companions merely excite eccentricities and inclinations in multi-planet systems, the lower mass companions would be preferentially removed as in standard planet scattering models \cite{Chatterjee2008}. 

Given the potentially competing processes governing planet formation and evolution for planets $>$100-day separations from stars in multi-star systems, it is challenging to predict {\it a priori} the number and distribution of small/rocky planets in the habitable zones of these stars. However, given the total fraction of Sun-like stars that are in multi-star systems, and that nearby Sun-like stars the most accessible targets for characterization of Earth-like planets by future missions (see Section \ref{sec:hwo}), it is compelling to try and answer the question.

A further interesting question is the potential habitability of Earth-like planets in multi-star systems. For a given system, there exists a maximum semi-major axis, $a_{crit}$, for each stellar component beyond which a planet orbiting that star is dynamically perturbed by the other components  \cite{Holman1999}. This value falls approximately linearly with the orbital eccentricity---the more eccentric the stellar orbit, the smaller the stable region around either star---and decreases with increasing mass ratio. For a multi-star system to be a suitable target for searching for Earth-like planets, the $a_{crit}$ values for the components stars would ideally be beyond the outer habitable zone limits for that star, implying a dynamically stable habitable zone. Simulations of the longevity of these stable zones as the binary orbit evolves indicated little change over $\sim$10My timescales; longer simulations would be needed to understand the long term impact on the prospect of stable habitable zones in multi-star systems \cite{Ballantyne2021}. The habitable zones in multi-star systems can also be shifted slightly outwards compared to single stars, depending strongly on the separation between the stars and more weakly on the different in effective temperatures of the stars\cite{Kaltenegger2017}.

\section{The SHERA Mission Concept}
\label{sec:SHERA}

Here we discuss the history, telescope, and orbit of the SHERA mission concept. We also present the diffractive pupil and other enabling aspects of the instrument, the target list, the survey design, and the science goals.

\subsection{History, Telescope, and Orbit}

The SHERA mission concept grew out of the TOLIMAN mission \cite{Tuthill2018,Bendek2018,Bendek2021,Tuthill2024,Tuthill2026}. TOLIMAN is a 12.5-cm telescope on a 16U CubeSat, led out of the University of Sydney\footnote{\url{https://toliman.space/}}. Planned for launch in 2027, TOLIMAN will perform microarcsecond-precision relative astrometry of $\alpha$ Centauri A and B for one year of operations. The development of TOLIMAN has advanced the technology needed for SHERA significantly, particularly the diffractive pupil design and the pipeline to extract the astrometric vectors, however the size of the TOLIMAN telescope restricts it the study of the $\alpha$ Cen system. Given the prevalence of nearby Sun-like stars in multi-star systems, the potential for small planets in these systems (see Section \ref{sec:binaries}), and the unsolved questions about the dominant planet formation and evolution processes in multi-star systems, there is significant scientific value in increasing the scale of TOLIMAN from one system to a larger survey, and a clear need for a concept like SHERA.

The SHERA instrument will be purpose-built to measure the relative separation of binary stars precisely enough to detect small exoplanets through their subtle gravitational influence on either stellar component. Achieving this science objective demands relative astrometric precision at the $\sim$$\mu$as-level over a three-year mission duration---corresponding to just six micropixels in the instrument’s focal plane. This would place SHERA’s astrometric precision at the forefront of space-based positional astronomy, targeting performance nearly 20 times more precise than Gaia’s current best-in-class global astrometry (see Section \ref{sec:stateoftheart}).

To achieve this demanding astrometric precision, the SHERA mission concept adopts a novel strategy: imaging bright, nearby, Sun-like stars with a closely separated (2--30$^{\prime\prime}$) Sun-like secondary as a reference frame, rather than relying on faint background stars \cite{Malbet2016}. As described in Section \ref{sec:relastro}, this approach overcomes the photon noise floor that typically limits the precision with which astrometry can be achieved with respect to faint reference stars, and opens up a heretofore unexploited discovery space around some of the nearest Sun-like stars. It also enables the use of a compact, small-aperture telescope.  The optical telescope will include a 22-cm aperture primary mirror, etched with a diffractive pupil (Section \ref{sec:pupil}), with a band-limiting filter providing 100-nm bandwidth centered on 550\,nm. The resulting narrow FOV simplifies the optical architecture to a lightweight two-mirror design with a band-limiting filter, and suppresses astrometric error terms that grow with angular separation. The high photon flux from the bright targets also supports high frame-rate (20 Hz) operation, mitigating image smear due to high frequency platform jitter, and enhancing temporal resolution for tracking any spacecraft dynamics. Higher frequency ($>$ 20 Hz) pointing jitter is a potential source of noise, smearing the PSF in potentially non-random ways, and the instrument requirement is to keep the smear below 0.05 arcseconds over a 50 ms time frame.

The SHERA spacecraft must reside in an orbit with a wide field of regard, accounting for a 90-degree Sun keep-out zone and a 60-degree Earth limb keep-out. The orbit must also limit the thermal gradients experienced by the instrument and avoid streaking caused by dense populations of spacecraft at low altitudes. To meet these requirements, a 650 km Sun-synchronous 6am-6pm orbit was selected, which is a cost-effective option to maximize the field of regard while maintaining a consistent thermal environment aligned with the solar terminator. Over the course of one Earth year, the instrument sees a full sweep of the sky, providing access to all targets. SHERA will perform a three-year survey of the inner planetary systems of nearby Sun-like stars (described in Section \ref{sec:survey}). The 14 stars targeted in the prime mission, and their orbital elements, are listed in Tables \ref{tab:targets} and \ref{tab:binaries}, respectively, and described in Section \ref{sec:targets}.

\subsection{The Diffractive Pupil and Other Enabling Aspects of the Instrument} 
\label{sec:pupil}

\begin{figure}
\begin{center}
\begin{tabular}{c}
\includegraphics[height=7cm]{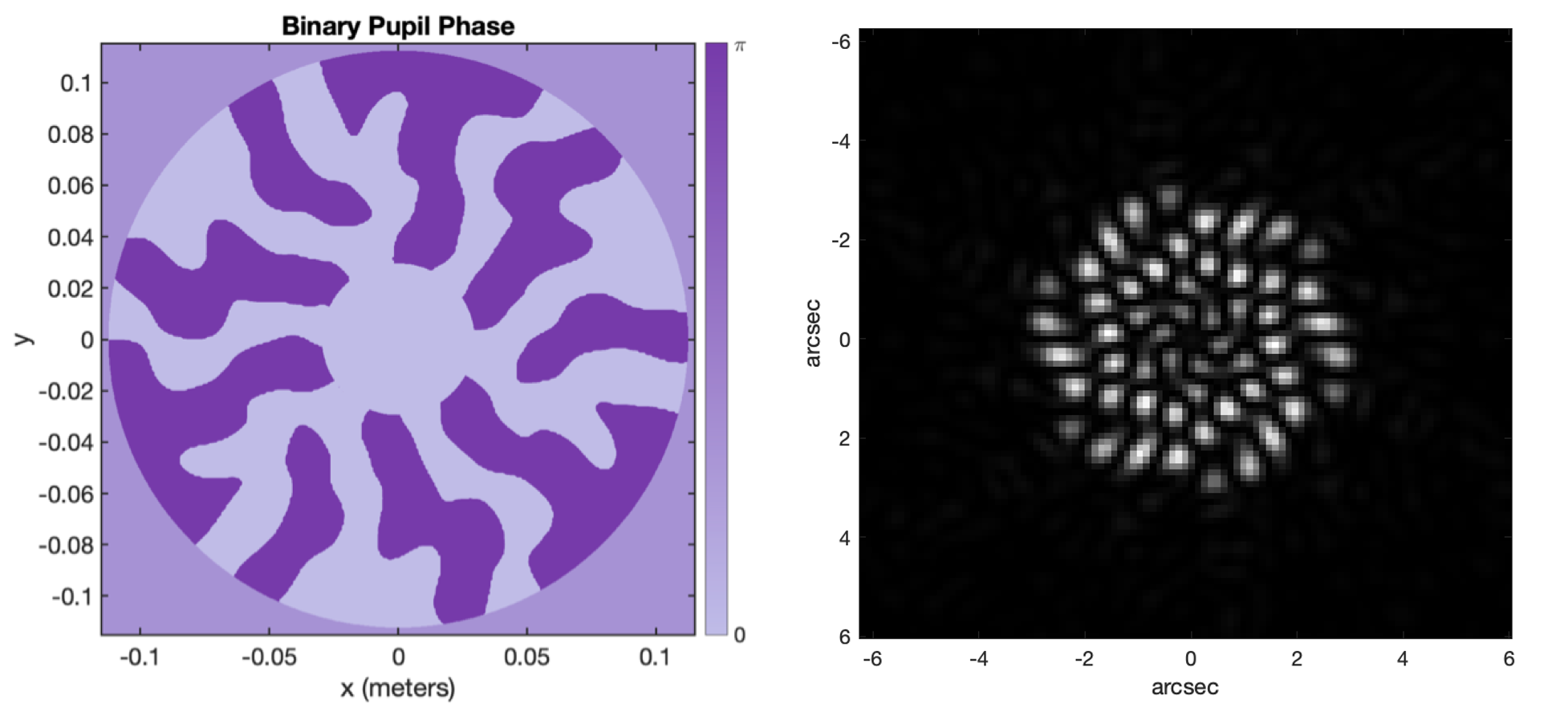}  
\\
\end{tabular}
\end{center}
\caption 
{\emph{Left}: The three-fold symmetry diffractive pupil phase pattern imprinted on the SHERA 22-cm primary mirror. \emph{Right}: Simulated stellar point spread function on the detector, sampled at $\simeq$0.123$^{\prime\prime}$.
\label{fig:pupil}} 
\end{figure}

At the core of the astrometry instrument is the diffractive pupil \cite{Bendek2021,Tuthill2024,Desdoigts2024}, which is electron-beam lithographed \cite{8088-2002} onto the telescope's primary mirror. The diffractive pupil is a binary phase mask ($0/\pi$ for spectral monitoring) with odd-fold circular symmetry (to break even-Zernike sign degeneracies), which is designed to be Fisher-information-optimal for marginal astrometric precision \cite{Desdoigts2024}. Figure~\ref{fig:pupil} shows a possible compliant diffractive pupil design, adapted from the TOLIMAN diffractive pupil. The final design will be determined by an ongoing detailed and principled optimization on the finalized SHERA optical design. The diffractive pupil's engineered phase variations, introduced in the optical pupil, serve as a metrological reference and fulfill three critical functions. First, the diffractive pupil enables image-plane wavefront sensing \cite{Desdoigts2024}, which allow SHERA to measure and track temporal variations in optical misalignment and aberration (e.g., focus, distortion, etc.) that, if not properly accounted for, strongly degrade astrometric performance. Second, the diffractive pupil is designed to spread starlight across multiple diffraction-limited Point Spread Function (PSF) features. This reduces astrometric error in the presence of variations in detector response and geometry, and relaxes the requirement for calibration of these detector variations more than an order-of-magnitude. Third, the diffractive-pupil-imposed phase preserves astrometric precision by maintaining intensity gradients in the PSF, ensuring that binary separations can be measured to within a small factor of the precision achievable with a classic Airy PSF. The PSF produced by the diffractive pupil is shown in the right panel of Figure~\ref{fig:pupil}. 

Since the DP serves as a metrological ruler, the intrinsic stability of the phase pattern it imposes is important. For this reason, it is fabricated on a low coefficient-of-thermal-expansion (`ultra-low expansion' or ULE) glass substrate. This design allows SHERA to relax the temperature stability requirements of the primary mirror to approximately $\sim$1K RMS, a level that is straightforward to achieve. Nevertheless, keeping the instrument temperature within prescribed bounds drives several other aspects of the mission design, including the selection of SHERA’s Sun-synchronous orbit, the implementation of a thermal skirt around the instrument, and an imposition of Sun-avoidance constraints on the telescope boresight, which in turn restrict when targets can be viewed. The expected  timescales of variability include: (i) the 30-minute observation window where the spacecraft is pointed at one target, (ii) the $\sim$98-minute low-Earth-orbit, (iii) seasonal variability (1--3 months) as the combination of visible targets changes throughout the year and results in different average angles of the spacecraft to the Sun over the course of several orbits, (iv) annual variability given the orbit of the Earth.

Beyond optical stability, SHERA’s astrometric precision relies on accurate knowledge of detector pixel locations and relative gains. Variations in these properties arise from detector manufacturing imperfections as well as non-uniform intra-pixel response \cite{Zhai2011}. These must be characterized extensively during ground calibration (to 0.1\% accuracy), and then continually refined in orbit using a data-driven self-calibration pipeline, adjusting flat-field and pixel position estimates \cite{Desdoigts2023}. This adaptive calibration scheme is needed to ensure that sensor evolution in the space environment does not degrade astrometric performance over the mission lifetime.

Additionally, since SHERA relies on the stability of diffracted PSFs for the binary pair, astrometric performance will depend on any changes in the effective wavelength of the incoming light and how well those changes are tracked. The accepted spectral band is defined by a band-limiting filter, and drifts in the filter’s dielectric coatings would produce PSF breathing in both stars, placing temperature stability requirements on the filter. In addition, temporal changes in stellar color (e.g., from photospheric activity) can alter the effective wavelength in a differential manner. To monitor and correct for these effects, SHERA will perform simultaneous spectroscopic measurements using a secondary set of diffractive features in the pupil that generate paired satellite spectra ($R$$\sim$500). These spectra allow the effective wavelength to be continuously tracked and separated from the astrometric signal.

Finally, for more than 90\% of the 3-year planned mission duration there is more than one SHERA prime mission system available for observation. The current plan is to cycle between the typically 3--5 systems that are visible at any given time, changing targets with each $\sim$98-minute SHERA orbit. This provides an opportunity to compare the diffractive pupil modeling performance between targets observed concurrently in time, and to identify any residual common-mode systematics across systems that may be due to incomplete capture and correction of the changing instrument state. We plan to monitor the evolving set of astrometric time series collectively to both improve the instrument forward modeling and detrend and remove common systematics.

The SHERA mission concept represents a highly integrated system of optical, thermal, and sensor engineering innovations, converging toward a single goal: enabling sub-$\mu$as relative astrometry between bright binary stars with a compact, space-based platform. Its diffractive pupil design enables a deep understanding of the interplay between instrument physics and astrometric measurement, and will advance the frontier of precision self-calibration of instrumental effects previously exemplified in precise exoplanet photometry in missions such as Kepler.

\subsection{Target List} 
\label{sec:targets}

In order to maximize the science return on the diffractive pupil imaging technology, SHERA will observe a target list of Sun-like stars in bright, nearby multi-star systems. Using the Washington Double Star catalog (WDS)\cite{Mason2001, Mason2023yCat....102026M} and SIMBAD, we created an initial target list of all binary stars within 50~pc with a primary magnitude brighter than V$<$7 mag, a secondary fainter by $<$3 magnitudes (i.e., down to 10$^{\rm th}$ magnitude), and a separation between 2--40$^{\prime\prime}$. The final list consisted of $\sim$200 individual stars. The projected angular separations on the sky are required to be wide enough to clearly separate the PSFs generated by the diffractive pupil for the range of apertures surveyed, but narrow enough to be comfortably sampled on the detector. The $\Delta$mag constraint ensured that the stars were not too widely separated in predicted flux levels.

After an iterative process of estimating the sensitivity of these stars to astrometric perturbations by planets---scaling for the aperture size and mission length---we identified the subset of the initial sample that were amenable to the detection of small, temperate exoplanets with a SHERA-like mission. Hotter stars, and cool, evolved stars with sufficiently bright companions, are rare by volume, and the nearest examples that satisfied our criteria were sufficiently distant that the $d^{-1}$ term in Eq. \ref{eq:astrometry} made them impracticable for efficient detection of planets with small apertures. The lower-mass M dwarfs were too faint, as only a single M dwarf over the entire sky is brighter than $m_V \simeq 7$. The sample was also restricted to systems for which the available data suggests they are two stars orbiting each other with no other known stellar companions within the FOV. We simulated the potential SHERA yield for a variety of scenarios using the full target list and a three-year mission profile. Given the steep increase in planet frequency below Neptune-mass planets ($\sim$17 $M_E$, e.g. \cite{Datillo2024}), we found that a narrow, deep survey yielded significantly more exoplanet detections and of more scientifically compelling planets that are otherwise inaccessible, than a wider, shallower survey sensitive only to larger planets. The optimal trade between survey depth per target and ensuring a number of targets with the statistical power to calculate meaningful statistics, was for a deep survey of the 14 targets that comprise the final SHERA prime mission sample. These 14 targets are listed in Table \ref{tab:targets}; their magnitudes lie between $0<V<7$, and they are all within $d\lesssim 17$\,pc. The sample is composed entirely of G- and K-type stars, all of which, given their projected on-sky separations of 5--32$^{\prime\prime}$, appear on the HWO potential target list \cite{Tuchow2025}, and 13 of the 14 are Tier 1 HWO targets \cite{Mamajek2024}. All distances were derived from Gaia Data Release 3 (DR3) \cite{Gaia2023AA...674A...1G} parallaxes, except in the case of the $\alpha$ Cen system\cite{Akeson2021}. The separation in 2031 ($\rho_{2031}$, the expected launch date of SHERA) was calculated using ephemerides calculated using the orbits referenced in Table \ref{tab:binaries}. 
Inner and outer limit habitable zones for planets on S-type orbits for each binary component were calculated using the {\it dihz} code from \cite{Eggl2020} and are listed in Table 1. In comparison to the single-star HZ limits, the binary star HZ limits typically differ by $<$0.5\%, with the largest effects for the outer HZ limits for $\alpha$ Cen B (moving outwards by 1.2\%) and $xi$ Boo B (moving outwards by 2.2\%).
%
%
%
All of the primary target stars are G- and K-type dwarfs, with a relatively narrow range of habitable zones ($\sim0.5-2.2$ au, orbital periods $\sim0.2-3.0$ yr), for which SHERA's astrometric survey would be most sensitive to planetary companions.

Retaining the same $V$magnitude and separation constraints, but loosening the distance limit to 50 pc, in Table \ref{tab:fulltargetlist} we list additional potential SHERA targets (16 systems, 32 stars). Double stars that had previously discovered to have resolved or unresolved higher order multiplicity are still omitted in this list (i.e. known triples, quadruples with separations of $<$40$^{\prime\prime}$). Given the range of apparent $V$ magnitudes and extended distances, the extended sample includes more luminous AFG-type main sequence evolved (subgiant, giant) stars, and the corresponding astrometric sensitivity is limited to massive planets ($>$20$M_{\oplus}$). 
The ORB6 \cite{Hartkopf2001} catalog of visual binary orbits\footnote{Regularly updated by R.~A. Matson et al. at \url{https://www.astro.gsu.edu/wds/orb6.html} } was consulted for orbits to calculate ephemerides during a prospective SHERA mission. Some of the systems are wide, slow-moving binaries for which orbits have not been published---for those, we examined WDS published data and Gaia DR3 astrometry and projected separations for epoch 2031 using linear or quadratic fits anchored to the precise separations measured by Gaia DR3 for epoch 2016.\\

\begin{table}
    \centering
    \footnotesize
    \begin{tabular}{|ccccccccccccc|}
    \hline
Name & $d$ & $\rho_{2031}$ & $V$ &  Ref. &  SpT & Ref. & log($L/L_{\odot}$) & Ref. & $r_{\rm HZ,in}$ & $r_{\rm HZ,out}$ & $\alpha$ & $M_p$\\
 & (pc) & ($^{\prime\prime}$) & (mag) & & & & (dex) & &
(au) & (au) & ($\mu$as) & ($M_{\oplus}$)\\
    \hline
$\alpha$ Cen A &  1.33 & 10.12 & 0.00 & \cite{Mermilliod1997} & G2V & \cite{Gray2006} & 0.180 & \cite{Soubiran2024} & 0.92 & 2.18 & 0.526 & 0.48\\
$\alpha$ Cen B &      &    & 1.35 & \cite{ESA1997} & K1V & \cite{Hoffleit1982} & -0.312 & \cite{Soubiran2024} & 0.53 & 1.25 & 0.526 & 0.71\\
\hline
61 Cyg A &  3.50 & 32.38 & 5.21 & \cite{Hauck1998} & K5V & \cite{Keenan1989} & -0.824 & \cite{Soubiran2024} & 0.29 & 0.69 & 1.11 & 3.8 \\
61 Cyg B &      &     & 6.04 & \cite{Hauck1998} & K7V & \cite{Keenan1989} & -1.013 & \cite{Soubiran2024} & 0.23 & 0.55 & 1.11 & 4.1\\
\hline
70 Oph A &  5.12 & 6.56 & 4.22 & \cite{ESA1997} & K0-V & \cite{Keenan1989} & -0.289 & \cite{GaiaDR3} & 0.54 & 1.27 & 0.638 & 3.0\\
70 Oph B &      &    & 6.06 & \cite{Eggenberger2008} & K4V  & \cite{Cowley1967} & -0.784 & \cite{Stassun2019} & 0.30 & 0.72 & 0.638 & 4.6\\
\hline
36 Oph A &  5.94 & 5.30 & 5.07 & \cite{ESA1997} & K1V & \cite{Torres2006} & -0.476 & \cite{GaiaDR3} & 0.43 & 1.02 & 0.563 & 3.7\\
36 Oph B &      &    & 5.11 & \cite{ESA1997} & K1V & \cite{Luck2017} & -0.476 & \cite{GaiaDR3} & 0.43 & 1.02 & 0.563 & 3.7\\
\hline
$\xi$ Boo A &  6.75 & 4.31 & 4.54 & \cite{ESA1997} & G7V & \cite{Gray2003} & -0.256 & \cite{GaiaDR3} & 0.56 & 1.32 & 1.028 & 6.8\\
$\xi$ Boo B &      &    & 6.98 & \cite{Mermilliod1997} & K5V & \cite{Abt1981} & -0.892 & \cite{GaiaDR3} & 0.27 & 0.65 & 1.028 & 9.7\\
\hline
p Eri A &  8.19 & 11.34 & 5.76 & \cite{Twarog1995} & K2V & \cite{Gray2006} & -0.478 & \cite{GaiaDR3} & 0.43 & 1.02 & 0.973 & 8.1\\
p Eri B &      &     & 5.88 & \cite{Twarog1995} & K2V & \cite{Gray2006} & -0.507 & \cite{GaiaDR3} & 0.42 & 0.99 & 0.973 & 8.3\\
\hline
HR 2667 &  17.1 & 21.94 & 5.56 & \cite{ESA1997} & G1.5V & \cite{Keenan1989} & 0.182 & \cite{GaiaDR3} & 0.92 & 2.18 & 0.682 & 6.8\\
HR 2668 &     &     & 6.83 & \cite{ESA1997} & K0.5V & \cite{Gray2006} & -0.302 & \cite{GaiaDR3} & 0.53 & 1.25 & 0.682 & 10.6\\
\hline
    \end{tabular}
    \caption{SHERA Binary Star Target List and Habitable Zones. Predicted separation $\rho_{2031}$ for epoch J2031.0 is calculated using the orbits in Table 2. $r_{\rm HZ,in}$ and $r_{\rm HZ,out}$ are the optimistic habitable zone limits\cite{Kopparapu2013}, scaled to account for the effects of the other companion star following \cite{Eggl2020}. $\alpha$ is the required integrated astrometric precision for each star, and $M_p$ is the planet mass in a one-year orbit that the precision would recover with a False Alarm Probability (FAP) of 0.001.
    \label{tab:targets}}
\end{table}

\begin{table}
    \centering
    \small
    \begin{tabular}{|cccccccccccc|}
    \hline
Name & $d$ & $\rho_{2031}$ & Ref. & $V$ &  Ref. &  SpT & Ref. & log($L/L_{\odot}$) & Ref. & $r_{\rm HZ,in}$ & $r_{\rm HZ,out}$ \\
 & (pc) & ($^{\prime\prime}$) & & (mag) & & & & (dex) & &
(au) & (au) \\
    \hline
$\gamma$ Vir A  & 12.02 & 4.01 & \cite{Scardia2007} & 3.47 & $^b$ & G1V    & \cite{Gray1989} & 0.669 & $^a$ & 1.6 & 3.8\\
$\gamma$ Vir B  &       &      &   & 3.52 & $^b$ & F0mF2V & \cite{Gray1989} & 0.658 & \cite{GaiaDR3}  & 1.6 & 3.8\\
\hline
$\psi^1$ Dra A  & 22.07 & 29.49 & \cite{Kisselev2009} & 4.57 & \cite{ESA1997} & F5IV-V & \cite{Gray2001} & 0.802 & \cite{Boyajian2012} & 1.9 & 4.5\\
$\psi^1$ Dra B  &       &       &   & 5.81 & \cite{ESA1997} & F8V    & \cite{Gray2001} & 0.318 & \cite{Stassun2019} & 1.1 & 2.6\\
\hline
$\iota$ Leo A  & 23.60 & 2.46 & \cite{Izmailov2025} & 4.03 & $^c$ & F1IV & \cite{Edwards1976} & 1.031  & $^a$ & 2.5 & 5.8\\
$\iota$ Leo B  &       &      &    & 6.68 & $^c$ & G3V  & \cite{Edwards1976} & -0.020 & $^a$ & 0.7 & 1.7\\
\hline
HR 7294 &  25.24 & 6.91 & \cite{Izmailov2025} & 6.57 & \cite{Mermilliod1997} & G2V & \cite{Gray2003} & 0.120 & \cite{GaiaDR3} & 0.9 & 2.0\\
HR 7293 &        &      &    & 6.75 & \cite{Mermilliod1997} & G3V & \cite{Gray2003} & 0.062 & \cite{GaiaDR3} & 0.8 & 1.9\\
\hline
17 Crt A &  28.15 & 9.66 & $^d$ & 5.58 & \cite{ESA1997} & F8IV-IV & \cite{Gray2006} & 0.557 & \cite{GaiaDR3} & 1.4 & 3.4\\
17 Crt B &        &      &    & 5.73 & \cite{ESA1997} & F9V     & \cite{Gray2006} & 0.517 & \cite{GaiaDR3} & 1.4 & 3.2\\
\hline
HR 9074 &  28.90 & 2.72 & \cite{Izmailov2025} & 6.47 & \cite{GaiaDR3} & F8 & \cite{Struve1955} & 0.280 & \cite{GaiaDR3} & 1.0 & 2.4\\
HR 9075 &        &      &    & 6.68 & \cite{GaiaDR3} & G1 & \cite{Struve1955} & 0.197 & \cite{GaiaDR3} & 0.9 & 2.2\\
\hline
$\alpha$ CVn A &  32.66 & 19.21 & $^e$ & 2.89 & \cite{ESA1997} & A0II-III & \cite{Gray2003} & 1.970 & \cite{Shultz2022} & 7.3 & 17.1\\
$\alpha$ CVn B &        &       &    & 5.61 & \cite{ESA1997} & F2V      & \cite{Gray2003} & 0.687 &  \cite{GaiaDR3} & 1.7 & 3.9\\
\hline
HR 6106 &  32.74 & 3.70 & $^f$ & 5.84 & \cite{Corbally1984} & F9IV & \cite{Gray2006} & 0.623 & \cite{GaiaDR3} & 1.5 & 3.6\\
HR 6105 &       &      &    & 6.63 & \cite{Corbally1984} & G0V  & \cite{Gray2006} & 0.300 & \cite{GaiaDR3} & 1.1 & 2.5\\
\hline
$\mu$ Vel A &  34.38 & 2.28 & \cite{Izmailov2025} & 2.71 & \cite{Kharchenko2009} & G6III & \cite{Gray2006} & 1.640 & \cite{Charbonnel2020} & 5.0 & 11.7\\
$\mu$ Vel B &        &      &    & 5.51 & \cite{Kharchenko2009} & G2V   & \cite{Edwards1976} & 0.770 & $^a$ & 1.8 & 4.3\\
\hline
$\gamma^2$ Del &  35.74 & 8.65 & \cite{Hale1994} & 4.27 & \cite{ESA1997} & K2IV & \cite{Slettebak1963} & 1.441 & \cite{GaiaDR3} & 3.9 & 9.3\\
$\gamma^1$ Del &        &      &    & 5.15 & \cite{ESA1997} & F8V  & \cite{Slettebak1963} & 0.973 & \cite{GaiaDR3} & 2.3 & 5.4\\
\hline
$\gamma^1$ Leo &  39.84 & 4.76 & \cite{Romanenko2014} & 2.33 & $^g$ & K1-III & \cite{Keenan1989} & 2.40 & \cite{Takeda2023} & 11.9 & 28.1\\
$\gamma^2$ Leo &        &      &    & 3.48 & $^g$ & K7IIIb & \cite{Keenan1989} & 1.80 & \cite{Takeda2023} &  6.0 & 14.1\\
\hline
$\epsilon$ Mon A &  39.85 & 12.08 & $^h$ & 4.39 & \cite{ESA1997} & A8V(n) & \cite{Gray2003} & 1.352 & \cite{Stassun2019} & 3.6 & 8.4\\
$\epsilon$ Mon B &        &       &    & 6.72 & \cite{ESA1997} & F6V    & \cite{Gray2003} & 0.448 & \cite{GaiaDR3} & 1.3 & 3.0\\
\hline
$\theta^1$ Ser &  40.93 & 22.74 & $^i$ & 4.62 & \cite{ESA1997} & A5V  & \cite{Gray1989} & 1.260 & \cite{Stassun2019} & 3.2 & 7.6\\
$\theta^2$ Ser &        &       &    & 4.98 & \cite{ESA1997} & A6Vn & \cite{Gray1989} & 1.126 & \cite{GaiaDR3} & 2.7 & 6.5\\
\hline
HD 218269 &  41.08 & 8.96 & $^j$ & 6.22 & \cite{Kharchenko2009} & F5V      & \cite{Gray2006} & 0.644 & \cite{GaiaDR3} & 1.6 & 3.7\\
HD 218268 &       &      &    & 6.99 & \cite{Kharchenko2009} & F5.5IV-V & \cite{Gray2006} & 0.321 & \cite{GaiaDR3} & 1.1 & 2.6\\
\hline
$\epsilon$ Dra A &  46.83 & 3.18 & \cite{Izmailov2025} & 3.88 & \cite{Kharchenko2009} & G7IIIb & \cite{Keenan1989} & 2.033 & \cite{GaiaDR3}  & 7.8 & 18.4\\
$\epsilon$ Dra B &        &      &    & 6.77 & \cite{Kharchenko2009} & F6     & \cite{Adams1935} & 0.525 & $^a$ & 1.4 & 3.2\\
\hline
107 Aqr A &  48.82 & 7.10 & $^k$ & 5.62 & \cite{Kharchenko2009} & A9IV & \cite{Abt1981} & 1.213 & \cite{Stassun2019} & 3.0 & 7.2\\
107 Aqr B &        &      &    & 6.53 & \cite{Kharchenko2009} & F2V  & \cite{Abt1981} & 0.645 & \cite{Stassun2019} & 1.6 & 3.7\\
\hline
    \end{tabular}
    \caption{SHERA Extended Binary Star Target List.  $\rho_{2031}$, $r_{\rm HZin}$, and $r_{\rm HZout}$ are as defined for Table \ref{tab:targets}. \footnotesize $^a$Estimates calculated based on the $V$ magnitudes and distances listed, and adopting BC$_V$ values from \cite{Pecaut2013} based on the spectral types listed. $^b$Calculated adopting $V_{AB}$ = 2.74, $\Delta V$ = 0.05. $^c$Calculated adopting $V_{AB}$ = 3.94, $\Delta V$ = 2.65. $^d$17 Crt is a wide binary with no ORB6 orbit. Separation between discovery epoch 1783 and Gaia DR3 (epoch 2016) varies approximately linear: $\rho$ = 9.585$^{\prime\prime}$ + 0.00494$^{\prime\prime}$ yr$^{-1}$($t_{yr}-2016$), $\theta$ = 210$^{\circ}$.102 - 0.0000814$^{\prime\prime}$ yr$^{-1}$ ($t_{yr}-2016$). $^e$$\alpha$ CVn is a wide binary with no ORB6 orbit. Separation is calculated using quadratic fit to separation data between 1940 and Gaia DR3. $^f$HR 6106/6105 is a wide binary with no ORB6 orbit. Separation between 1980 and Gaia DR3 varies approximately linearly: $\rho$ = 4.070$^{\prime\prime}$\ - 0.0248$^{\prime\prime}$ yr$^{-1}$ ($t_{yr}-2016$). $^g$Calculated adopting $V_{AB}=2.01$ \cite{ESA1997}, $\Delta V =1.15$ \cite{Rakos1982}. $^h$$\epsilon$ Mon is a wide binary with no ORB6 orbit. Separation is calculated using quadratic fit to separation data between 1985 and Gaia DR3. $^i$$\theta$ Ser is a wide binary with no ORB6 orbit. Separation between 2003 and 2018 was approximately linear: $\rho$ = 22.457$^{\prime\prime}$ + 0.01887 $^{\prime\prime}$\,yr$^{-1}$ ($t_{yr}-2016$). $^j$HD 218269/218268 is a slow binary with no ORB6 orbit. Separation data between 1975 and 2016 are approximately linear: $\rho$ = 8.895$^{\prime\prime}$ + 0.00405 $^{\prime\prime}$\,yr$^{-1}$ ($t_{yr}-2016$). $^k$107 Aqr is a slow binary with no ORB6 orbit. Separation data between 1980 and 2018 are approximately linear: $\rho$ = 6.983$^{\prime\prime}$ + 0.008083$^{\prime\prime}$\,yr$^{-1}$ ($t_{yr}-2016$).}
    \label{tab:fulltargetlist}
\end{table}

The knowledge of the stellar parameters and orbits of these systems was heterogeneous, ranging from extremely well-studied (e.g., $\alpha$ Cen, 61 Cyg) to much less constrained (e.g., HR 2667/2668). However, an in-depth knowledge of the orbital parameters in each system is necessary to achieve the science goals of the SHERA mission. A thorough literature search was conducted for each system, and the most recent orbits were found and used to predict ephemerides, ensuring that the systems still satisfy our angular separation criteria over the next decade. A summary of the state-of-the-art orbits for the seven selected systems is summarized in Table \ref{tab:binaries}. The orbits for $\alpha$ Cen, 61 Cyg, and $\xi$ Boo were found to be sufficient for initial calculations regarding the stability of planets in the habitable zones of the stars, while the orbits for the others were either poorly constrained or had no quoted uncertainties. \cite{Li2026arXiv260320044L} combines recent and historical astrometry with new precise radial velocity observations to produce a greatly improved solution for the 70 Oph system. Preliminary orbits have been also calculated for 36 Oph, p Eri, and HR 2667/2668 using archival astrometry and radial velocity measurements. Stable orbital zone outer radii for S-type orbits ($a_{crit}$) are calculated\cite{Holman1999} using the orbital elements listed in Table \ref{tab:binaries}.
The $a_{crit}$ values range from $\sim$0.8--41 au, and by comparison with the habitable zone limits calculated in Table \ref{tab:targets}, we infer that the conservative habitable zones for these stars all lie within their estimated Holman-Wiegert stable S-type orbit zones, with the exception of 36 Oph. For this system, the estimated $a_{crit}$ value for both stars (0.83\,au) is slightly smaller than the estimated outer habitable zone limit (1.02\,au); for both stars more than 50\% of the habitable zone is within the theoretical stability limit.

\begin{table}
    \centering
    \small
    \begin{tabular}{|ccccccccccc|}
    \hline
Name & Ref. & $P$ & $a$ & $i$ & $\Omega$ & $T$ & $e$ & 
$\omega$ & {\bf $a_{crit}$} & {\bf $P_{crit}$} \\
 &  & (yr) & ($^{\prime\prime}$) & (deg) & (deg) & (yr) &  & (deg) & (au) & (yr) \\
\hline
$\alpha$ Cen & \cite{Akeson2021} & 79.76 & 17.5 & 79.243 & 205.07 & 1955.56 & 0.5195 & 231.52 & 2.74 & 4.30\\
       &  & $\pm$0.02 & $\pm$0.01 & $\pm$0.009 & $\pm$0.03 & $\pm$0.02 & $\pm$0.0002 & $\pm$0.03 & 2.49 & 4.05\\
\hline
61 Cyg & \cite{Giovinazzi2025} & 707.1 & 24.81 & 53.0 & 355.7 & 2396 & 0.442 & 151 & 11.9 & 50.0\\
       &  & $\pm$0.5 & $\pm$0.05 & $\pm$0.1 & $\pm$0.3 & $\pm$3 & $\pm$0.006 & $\pm$1 & 11.4 & 48.6\\
\hline
70 Oph & \cite{Li2026arXiv260320044L} & 88.13 & 4.5433 & 121.14 & 121.67 & 2072.17 & 0.5002 & 193.15 & 2.88 & 5.21\\
       &  & $\pm$0.01 & $\pm$0.0009 &  $\pm$0.03 &  $\pm$0.03 &   $\pm$0.02 & $\pm$0.0002 &  $\pm$0.06 & 2.59 & 4.86\\
\hline
36 Oph & $^a$ & 520     & 12.265     & 100.58    & 92.4     & 2178.78    & 0.8858      & 89.90     & 0.83 & 0.83\\
       &      & $\pm$23 & $\pm$0.369 & $\pm$0.31 & $\pm$2.9 & $\pm$12.92 & $\pm$0.0059 & $\pm$1.15 & 0.83 & 0.83\\ 
\hline
$\xi$ Boo & \cite{Izmailov2025} & 152.46 & 4.920 & 140.54 & 167.9 & 2061.90 & 0.5141 & 24.0 & 4.16 & 8.67\\
          &  &  $\pm$0.07 & $\pm$0.003 &  $\pm$0.07 &  $\pm$0.2 & $\pm$0.09 & $\pm$0.0005 & $\pm$0.3 & 3.39 & 7.62\\
\hline 
p Eri & $^a$ & 364.4 & 7.221 & 139.40 & 172.92 & 2180.27 & 0.6244 & 336.9 & 5.02 & 12.5\\
      & & $\pm$2.1 & $\pm$0.017 & $\pm$0.46 & $\pm$0.95 & $\pm$1.69 & $\pm$0.0062 & $\pm$1.3 & 4.78 & 12.1\\
\hline 
HR 2667 & $^b$ & 4354 & 18.313 & 76.59 & 138.87 & 4227.5 & 0.667 & 133.3 & 23.2 & 114\\
        &  & $\pm$111 & $\pm$0.192 & $\pm$0.64 & $\pm$0.83 & $\pm$51.6 & $\pm$0.022 & $\pm$0.6 & 21.3 & 109\\
        \hline
    \end{tabular}
    \caption{SHERA Binary Star Target List Orbits and Stability Limits. The orbital elements listed are rounded values in standard format of ORB6 \cite{Hartkopf2001}; full orbital elements and uncertainties can be found in the references listed. $a_{crit}$ and $P_{crit}$ are the approximate maximum semi-major axes and orbital periods for stable S-type orbits around the binary components. $^a$Giovinazzi et al. (in prep). $^b$ Clark et al. (in prep).}
    \label{tab:binaries}
\end{table}

\subsection{SHERA Survey Design}
\label{sec:survey}


\subsubsection{Mission Design and Simulated Data}
\label{sec:missiondesign}

The SHERA concept of operations (ConOps) is optimized for high-cadence, high-precision AB relative astrometry of the mission's sample of prioritized binary star targets. SHERA operates in a Sun-synchronous low-Earth orbit, and maintains a strict Sun-avoidance angle of $\sim90^{\circ}$, constraining instantaneous target accessibility while enabling thermally stable and repeatable observing conditions. At any given time, systems will be scheduled for observation based on visibility windows and spacecraft attitude constraints. Target observations are conducted during continuous viewing arcs constrained by Earth occultation, Sun avoidance, and attitude limits. Given the $\sim$98-minute orbit, targets will be observed with typical uninterrupted observation blocks of $\sim$30 minutes at 20\,Hz, with up to three observation blocks scheduled per orbit.

Over its nominal three-year mission lifetime, SHERA will acquire a densely sampled time series of relative astrometric measurements for each binary system, maximizing phase coverage through optimized scheduling across the available visibility windows. The resulting datasets will have sensitivity by the end of the prime mission to support detection of the small perturbations induced by terrestrial-mass (0.4--4$M_{\oplus}$) planets. Signals are initially detected via periodogram-based analysis, then characterized by nonlinear orbital fitting to constrain planetary masses and orbital elements. The performance modeling of representative systems, such as $\alpha$ Cen AB, uses ConOps simulations of relative astrometric time series with an instrument error budget that predicts a single-measurement precision of $\simeq$\,4 $\mu$as per 30-minute observation block for the $\alpha$ Cen system. This model accounts for contributions from astrophysical noise sources (e.g. modeled intrinsic astrometric jitter of $\alpha$ Cen AB, optical contamination from known background field stars), noise from unmodeled instrumental variation (including the optics and detector subsystems), as well as standard detection noise terms (including photon noise, detector read noise etc.). Note that within a 30-minute observation block, the $\alpha$ Cen orbital motion as well as the AB differential aberration due to the spacecraft orbital motion will be significantly larger than the $4$\,$\mu$as noise floor.

\begin{figure}
\begin{center}
\begin{tabular}{c}
\includegraphics[height=5.8cm]{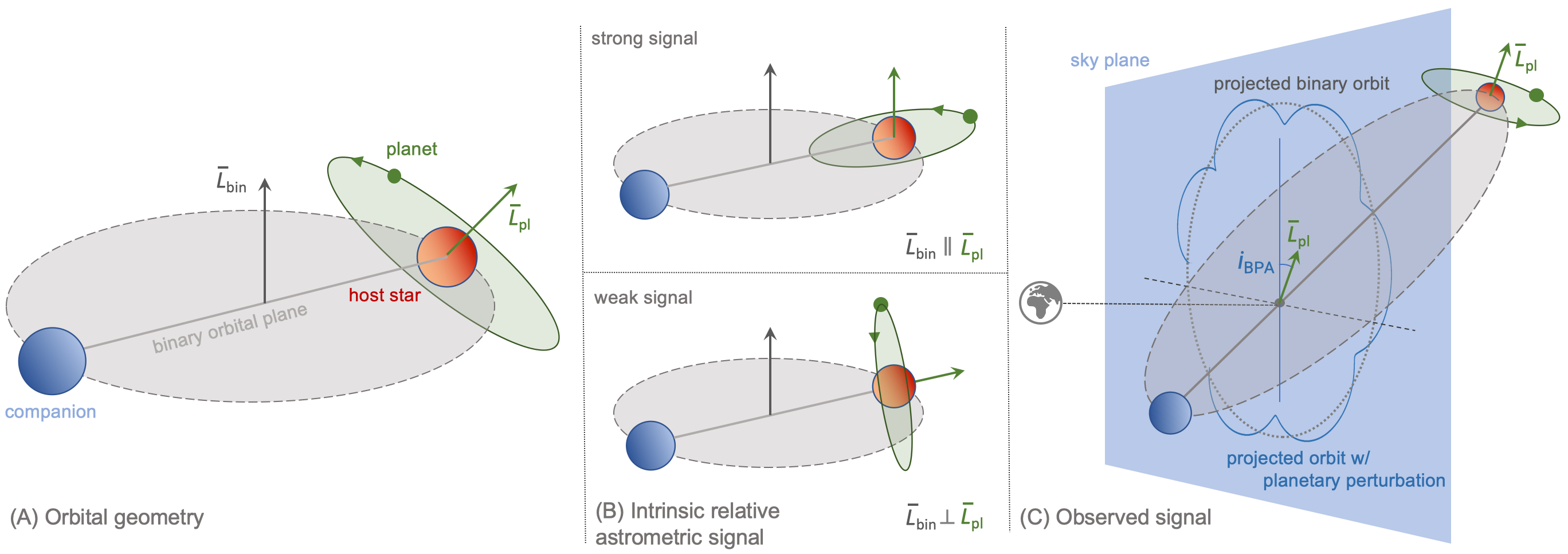}
\end{tabular}
\end{center}
\caption 
{(A) An illustration of the orbital geometry of a binary system (or binary component of a multi-star system) where one star (labeled `host star') has an orbiting planet (`b') with an angular momentum vector ($L_{\rm pl}$) at an arbitrary angle to the angular momentum vector of the binary ($L_{\rm bin}$). (B) The middle panel shows two configurations of the system: in the upper half, $L_{\rm bin}$ and $L_{\rm pl}$ are parallel, meaning that the planet is orbiting in the plane of the binary. This produces the strongest projected relative astrometry signal change for an observer viewing the binary system face-on. The lower half shows the case where $L_{\rm bin}$ and $L_{\rm pl}$ are perpendicular. The lowest projected relative astrometry signal will result when $L_{\rm pl}$ is perpendicular to $L_{\rm bin}$. (C) The final observed relative astrometric signal, projected onto a sky plane orthogonal to the observer. The original (unperturbed) binary orbit is shown as a grey dashed line, at an angle $i_{\rm BPA}$ to the plane of the sky, the binary orbit projected onto the sky plane is shown as a grey dotted line, and the projected orbit including perturbation from a planet is shown as a solid blue line (noting that the amplitude of the perturbations has been amplified for illustrative purposes). The observed planetary signal provides a minimum mass for the orbiting planet, analogous to the minimum mass that is measured for radial velocity observations.
\label{fig:orbit_orientation}}
\end{figure}

Since SHERA measures binary separation with much more precision than position angle (PA), the measurements are predominately one-dimensional. This results in a degeneracy between the orientation of the planet's orbital plane and the planet's mass, similar to the $M_p \sin(i)$ degeneracy for RV planets. In the case of one-dimensional astrometry, the relevant angle is the mutual inclination between the planet's orbital plane and a reference plane that is perpendicular to the on-sky vector connecting the two stars. We refer to this angle as $i_{\rm BPA}$ since it is derived from the PA of the binary companion, and the angle is illustrated in Figure \ref{fig:orbit_orientation}. Much the way a planet with an inclination of 0$^\circ$ to the plane of the sky will produce no RV signal, a planet with $i_{\rm BPA} = 0^\circ$ or $180^\circ$ will induce only a minimal detectable change in separation between the two stars. As a result, our orbit fit most precisely constrains $M_p \sin(i_{\rm BPA})$. For simplicity, the fitting is not performed in $M_p \sin(i_{\rm BPA})$ directly, but rather in coplanar mass: the mass the planet would have if the binary orbital planet and the planet’s orbital plane were the same. The coplanar mass is similar (but not identical) to $M_p \sin(i_{\rm BPA})$, but has a more direct physical meaning. For example, for the $\alpha$ Cen binary system, a coplanar mass of 1.000 $M_\oplus$ corresponds to an $M_p \sin(i_{\rm BPA})$ of 0.992 $M_\oplus$, assuming an instantaneous PA of 107.7$^\circ$.

\subsubsection{SHERA Completeness}

We begin with an analytic estimate of SHERA's sensitivity to planets as a function of stellar parameters, planet properties, observational strategy, and errors, comparing the expected astrometric reflex motion of the planet-hosting star to the expected SHERA astrometric uncertainty.

The amplitude of the astrometric reflex motion (in arcseconds) is given by

\[
a_{\rm reflex} = P^{2/3} \frac{M_p}{(M_1 + M_p)^{2/3}d}
\]

\noindent where $P$ is the orbital period in years, $M_1$ is the mass of the host star, $M_p$ is the mass of the planet (both in solar masses), and $d$ is the distance to the system in parsecs. This simplification is valid when the planet's eccentricity is 0 and the planetary orbit is in the plane of the sky.

We consider a planet detectable if the periodic signal reaches a False Alarm Probability (FAP) of 0.001, which corresponds to a $\sim$4-sigma detection of the astrometric signal, where he noise term, $\sigma$, is the single-epoch error divided by the square root of the total number of epochs.

Finally, we include an additional term to account for the decreased sensitivity to wider-separation planets where the observing baseline ($T$, nominally 3 years) is less than the planet's orbital period. By expanding the sinusoidal motion (under the assumption that baseline is much shorter than the orbital period), we derive an additional factor of ${\left ( \frac{T}{P} \right )} ^2$ which is applied to orbital periods longer than the baseline.

\begin{figure}
\begin{center}
\begin{tabular}{c}
\includegraphics[height=9cm]{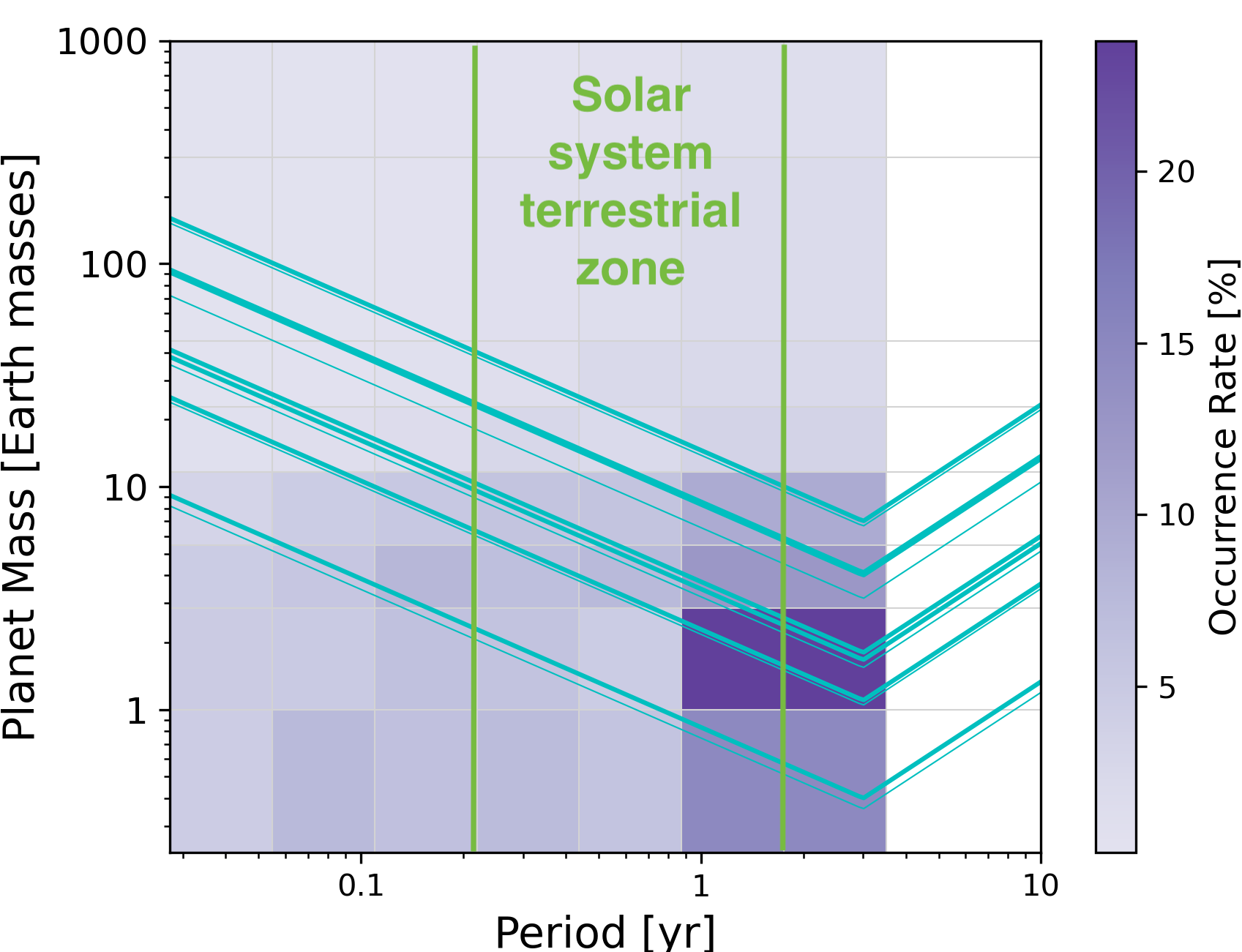}
\end{tabular}
\end{center}
\caption 
{Given SHERA's expected astrometric precision, the cyan lines show the 4$\sigma$ analytic mass sensitivity limits achievable for a survey of the 14 stars over a three-year mission duration.
\label{fig:sensitivity}}
\end{figure}

We show the sensitivity predictions for our fourteen targets in Figure \ref{fig:sensitivity}, ranging from 0.4\,$M_{\oplus}$ around $\alpha$ Cen, 2--3\,$M_{\oplus}$ around 36 Oph, 70 Oph, and 61 Cyg, and 4--10\,$M_{\oplus}$ around $\xi$ Boo, p Eri, and HR 2667/2668 for planets in 3-year orbits.

\subsubsection{Orbit Fitting}
\label{sec:orbitfitting}

We validate the analytic expression of SHERA sensitivity by performing a series of orbit fits to simulated planets using our custom orbit fitting pipeline. For these initial simulations, we perform a simplified fit, assuming that the orbital motion of the binary and spacecraft have been accounted for, and that the astrometric residuals contain only the putative planet signal/s. Future versions of the pipeline, described in more detail in Roberson et al. in prep, will include the full suite of the binary orbital motion, LEO orbital aberration, differential annual aberration, and differential annual parallax, accounting for the position and movement of the spacecraft relative to the solar system barycenter.

\begin{figure}
\begin{center}
\begin{tabular}{c}
\includegraphics[height=6cm]{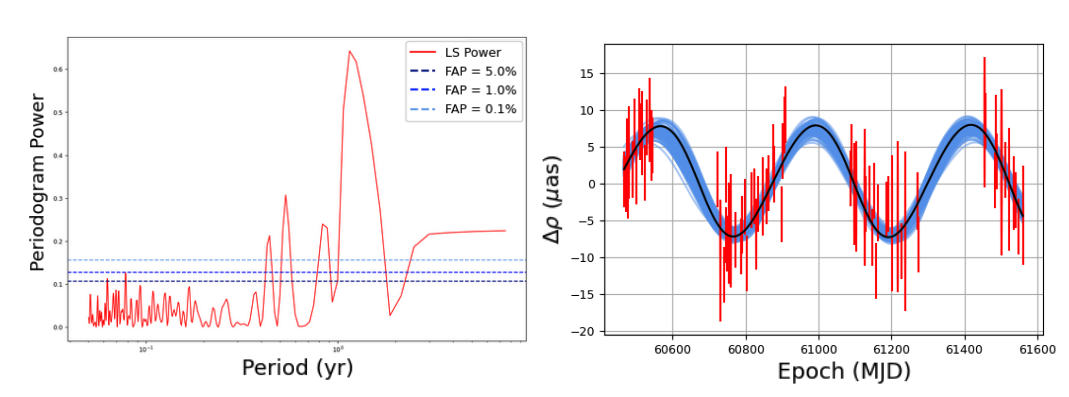}
\end{tabular}
\end{center}
\caption 
{With our custom pipeline we demonstrate accurate characterization of habitable zone Earth-mass planets from simulated SHERA data. In this example, a $3 \mearth$ planet in a 1.2-year orbit around $\alpha$ Cen A is clearly detected in the periodogram search (left). An MCMC fit to the data (right) recovers the simulated orbital period within 2\%, and coplanar mass to within 8.5\%.
\label{fig:detection_characterization}}
\end{figure}

Since relative astrometry is one-dimensional (analogously to RV measurements), the pipeline is modeled after RV methods used by the California Legacy Survey \cite{Rosenthal2021}, and is described in depth in Roberson et al. (in prep). As with the \texttt{RVsearch} package developed for the California Legacy Survey, we separate planet detection from orbit fitting, beginning with a periodogram search for signals in the data. Any signal above a set False Alarm Probability threshold is then characterized with a Markov Chain Monte Carlo (MCMC) fit to the orbital parameters of the planet, including planet mass. The outputs of the periodogram and MCMC analyses for one of these simulated planets are shown in Figure \ref{fig:detection_characterization}.

We simulated 10,000 SHERA astrometric time series with single planets orbiting $\alpha$ Cen A, and attempted to recover their signals using our pipeline. These planets were generated with random masses and periods drawn from log uniform distributions, with periods from 0.2 to 6.5 years, and masses from 0.1 to $3 M_\oplus$. An injected planet is considered recovered if there is a significant peak in the periodogram, which is set to search for periods up to 7.5 years, and the resulting estimate of period accurately recovers the injected value to within $P^2/2T$ of the injected signal; for a 365-day signal in a three-year mission, this would correspond to $\pm$61 days.

We find good agreement between our preliminary periodogram analysis and the analytic expectation of SHERA completeness. For example, the analytic expression predicts SHERA is sensitive to planets more massive than $0.4 M_\oplus$ with periods of 3 years. We recover 92\% of the simulated planets down to $0.4 M_\oplus$ at this period, and 95\% of the simulated planets down to $0.45 M_\oplus$. The majority of the non-detections are at masses only slightly greater than $0.85 M_\oplus$, as expected. The analytic expression predicts completeness to slightly lower masses than we see in our injection/recovery tests, which is likely due to the periodic gaps in the observing window. The analytic prediction assumes that we are able to see the full astrometric deflection of the host star during the mission, which may not be the case for certain orbital phases, as their maxima and minima may not be contained within the observing windows. We recover over 90\% of the simulated planets more massive than $1.2 M_\oplus$ with periods near 1 year.

Signals identified by the periodogram analysis are then characterized with a simplified initial MCMC orbit fitting procedure that fits for the planetary orbit only, measuring full posteriors for coplanar mass, eccentricity, orbital phase, and period (see Roberson et al. in prep for more detail). Remaining false positives (signals found by the periodogram analysis that are not associated with an injected signal) are identified at this stage from the MCMC posteriors, as we require a planet to have a False Alarm Probability of 0.001 to be considered detected. For these initial MCMC tests, we do not yet constrain the binary orbit, as we are fitting only to the difference in binary separation caused by the planet. Although we expect minimal binary orbital motion over the three years, the expected amplitude of the motion is high enough that we will need to simultaneously model for it. Our final orbit-fitting methodology will include priors on the binary orbit and fit for the full binary orbital elements.

Our planet detection and characterization pipeline also demonstrates that we are able to extract multiple planets from a single dataset. We run an additional suite of injection/recovery tests with 10,000 two-planet systems, which result in similar planet recovery rates to single-planet systems; we successfully recover $\sim$68\% of planets above 0.85 $M_\oplus$ with periods near 1 year, compared to $\sim$75\% for our single planet tests. This drop in sensitivity is partially driven by cases in the simulated datasets where two planets have similar periods. In particular, if the two periods are within a single periodogram bin, our analysis will typically only initially identify a single planet; the initial analysis does find both planets in $\sim$24\% of the simulated systems. In some cases it is possible to identify the presence of a second planet from the residuals to the MCMC fit, if the time baseline is long enough compared to the orbital period (see the $P^2/2T$ tolerance window mentioned above). Multi-planet systems with more distinct periods (for example, Venus at 225 days, Earth at 365 days, Mars at 687 days) are detected much more cleanly.

\subsection{Example Science Enabled by SHERA}
\label{sec:objectives}

SHERA will perform a dedicated survey of fourteen bright, nearby, Sun-like stars in multi-star systems, described in Section \ref{sec:targets}. As described earlier, given the steep increase in frequency with decreasing planet size (Figure \ref{fig:sensitivity}), a narrow (number of targets), but deep (planet mass sensitivity) survey produces a significantly higher planet yield in a given mission duration than a shallower survey over many more targets, and accesses scientifically compelling targets that are inaccessible with other techniques. With this survey, SHERA would enable a number of important science investigations. Here we describe three potential science cases, including: 1) searching for the nearest Earth-like planets; 2) measuring the occurrence rate of planets across the solar system terrestrial zone in multi-star systems; and 3) characterizing the three-dimensional architectures of planet-hosting binaries.

\subsubsection{Finding the Nearest Earth-like Planets}
\label{sec:SO1}

For the closest eight stars in the SHERA target list, shown in Table \ref{tab:targets} and Figure \ref{fig:astrometryRV}, SHERA will have sensitivity to planets down to 0.4--4\,$M_{\oplus}$ in their habitable zones, as shown in Figure \ref{fig:sensitivity}. Extending the SAG13 occurrence rates \cite{SAG13} to 3-year orbital periods, we expect to discover 4$\pm$2 small ($<$4\,$M_{\oplus}$), habitable zone planets, assuming the same frequency of planets around single stars (but see Section \ref{sec:SO2} below). Given that there are currently no robust confirmations of rocky planets in the habitable zones of Sun-like stars, any discoveries in this parameter space would be exceptionally interesting. As discussed in Section \ref{sec:relastro}, relative astrometry with SHERA provides evidence that one of the stars in a given binary system has a planet, but on its own does not uniquely identify which star is the host. Depending on the planet's mass, future radial velocity capabilities (Section \ref{sec:stateoftheart}) may be able to ascertain which of the stars hosts the planet (Section \ref{sec:SO3}), and further unveil the three-dimensional architecture of the system. Future direct imaging capabilities such as \HWO\ will be able to quickly determine which star the planet is orbiting. Indeed, any rocky, habitable zone planet candidates found in these nearby systems of bright, Sun-like stars would immediately become first-light targets for \HWO\ and other future imaging capabilities (see Section \ref{sec:hwo}).

\subsubsection{Measuring the Occurrence Rate of Planets in Multi-Star Systems}
\label{sec:SO2}

The primary set of targets most amenable to searching for small, habitable zone planets described above all (somewhat unfortunately but seemingly coincidentally) occupy the same 90$^{\circ}$ of right ascension. This means that there is additional observing time available when these targets are not available from the Earth. SHERA would use this time to perform a wider survey of binary systems, adding the six additional targets shown in Table \ref{tab:targets} and Figure \ref{fig:astrometryRV}. Since these targets are more distant and/or fainter than the primary set of targets, SHERA would not be as sensitive to the smallest planets in these systems, but would still typically be able to detect planets smaller than 10 $M_{\oplus}$ (Figure \ref{fig:sensitivity}). With the larger sample of 14 targets, SHERA would be able to test whether the single star SAG13 occurrence rates apply across the solar system terrestrial zone (SSTZ). The SSTZ is defined as the orbital period range 80--640 days, which roughly encompasses the orbits of the rocky planets in our solar system (see the green vertical lines in Figure \ref{fig:sensitivity}. Correcting the SAG13 statistics for unresolved binaries \cite{Bergsten2026}, we estimate the survey yield for two assumptions: (i) no suppression compared to single stars, resulting in $7.1\pm2.1$ planets; and (ii) that the suppression found in \cite{Moe2021} for shorter period planets applies to planets across the SSTZ, resulting in $3.7\pm1.9$ planets. If fewer than 2 planets are discovered, SHERA would rule out the SAG13 single star occurrence rates with a $>99.9$\% confidence ($>$3$\sigma$). This would provide strong evidence that planets are suppressed across the SSTZ of multi-star systems, extending the previous results that they are suppressed closer in (Section \ref{sec:binaries}). We will also be able to place an upper limit on the occurrence rate of more distant giant planets with periods beyond $\sim$6 years, which would appear as residual accelerations in the SHERA astrometry.

\subsubsection{Characterizing the 3D Architecture of Multi-Star Systems with Planets}
\label{sec:SO3}

The three-dimensional architectures of planet-hosting multi-star systems offer a unique laboratory for testing how planetary systems form and evolve. In particular, the distribution of angular momentum across the various components in these systems, whether aligned or mis-aligned, serves as a fossil record of the system’s past, tracing historical formation and evolutionary processes, such as migration within a disk \cite{Dupuy2022}, close planetary encounters \cite{Reynolds2024}, or significant gravitational influences from the secondary star \cite{Gerbig2024}. The three-dimensional architectures of hierarchical multi-star systems reveal which dynamical processes dominate throughout system lifetimes. 

Besides the small planets targeted in Section \ref{sec:SO1}, SHERA will also be sensitive to giant planets in the systems on a wider range of orbits (see Figure \ref{fig:sensitivity}), both interior and exterior to the solar system terrestrial zone. Although these planets may be accessible to other detection techniques, such as radial velocity, the relative astrometry measurements obtained by SHERA provide a 1D measurement in the plane of the sky (see, e.g. Figure \ref{fig:orbit_orientation}). Radial velocity measurements similarly provide an 1D measurement, although perpendicular to the plane of the sky. These orthogonal measurements can then be combined with stellar inclinations, constrained by stellar rotation periods and projected stellar rotational velocities ($v$ sin $i$) to obtain three-dimensional reconstructions of each orbit in the system. For planets discovered by SHERA that are accessible to the next generation of RV characterization (typically $>$5\,$M_{\oplus}$), the combination of astrometry from SHERA and ground-based RVs would provide the most precisely characterized 3D geometries yet for these hierarchical systems, allowing us to infer the importance of formation versus evolution in setting observed system states.

\subsection{Measurement requirements}

In order to reach the planet mass sensitivities at the orbital periods outlined in Section \ref{sec:SO1}, SHERA measures the time-varying angular separation of nearby binary stars. The fundamental observable is the scalar separation between the two stellar components, measured repeatedly over a three-year mission. The precision of this differential angular measurement $\theta_{AB}$ is fundamentally set by the diffraction limit and photon statistics, scaling approximately as $\delta\theta_{AB}\sim(\lambda_{eff}/D)\sqrt{(1/SNR_{A}^{2}+1/SNR_{B}^{2}})$, where $D$  is the telescope aperture, $\lambda_{eff}$ is the effective wavelength, and $SNR_{(A,B)}$ are the signal-to-noise ratios of the two stellar components. Because the measurement is differential, the companion star itself serves as a bright, co-moving astrometric reference, enabling substantially higher precision than conventional wide-field relative astrometry based on faint background stars\cite{Malbet2016}. The trade-off to this increase in precision is the introduction of an inherent degeneracy as to which component of the binary is the true host star. To achieve the required mass precision of 33\%, using simulations with the MARA pipeline (\ref{sec:orbitfitting}) we allow a 25\% contribution from the signal-to-noise of the detection, corresponding to a False Alarm Probability (FAP) threshold of $\le$0.001, and a 22\% contribution from other sources, such as the uncertainty on the mass of the host star. By selecting near-equal mass binaries, we minimize the contribution to the final planet mass uncertainty from the degeneracy of which component is the host star. To detect planets at the required masses with a FAP of $\le$0.001, the total integrated astrometric precision required ranges from 0.52--1.1 $\mu$as per target (\ref{tab:binaries}). In order to achieve the required orbital period coverage and precision, our simulations further indicate we require observations spanning at least one year per target. Finally, the occurrence rate analysis in Section \ref{sec:SO2} requires a sufficiently large sample size, with sufficient sensitivity to planets across 0.3--1.5 au, such that a null result (SHERA finding 0 planets) rules out intermediate-separation binaries having the same planet frequency across this parameter space as single stars at $\ge$3$\sigma$. To reach the required mass sensitivities with a FAP of $\le$0.001 and a corresponding planet mass uncertainty of $\sim$33\%, SHERA’s integrated astrometric precision requirements at a fixed orbital period of 1-year range from 0.52--1.1 $\mu$as across the 14 targets, listed in Table \ref{tab:targets}.

By jointly estimating the binary separation and instrument state with the diffractive pupil, SHERA relaxes many of the stringent engineering requirements that would otherwise be imposed on a $\mu$as-level mission. Rather than relying on extreme stability, SHERA’s time-varying instrumental effects are measured and incorporated into the astrometric solution. The ASTERIA mission noted correlated noise introduced by, for instance, pointing jitter that limited their precision on short ($<$ days) timescales \cite{Krishna2021, Seager2021}. A key feature of the SHERA measurement architecture is that the dominant residual instrumental errors are expected to be largely uncorrelated between observations, since systematic drifts in the instrument state are captured by the re-estimation and update of the instrument state with each observation. Consequently, instrumental uncertainties are expected to average down approximately as $N^{(-1/2)}$ over repeated observations and reach the required $\le$1 $\mu$as-level. Remaining correlated signals, such as stellar astrometric jitter or long-timescale astrophysical variability, are treated separately (see \cite{Meunier2022}) within the mission performance model and are not expected to follow this statistical averaging.

Finally, retrieving accurate planetary masses and orbital periods requires precise knowledge of the binary orbits and distances. The SHERA prime mission targets are all bright, well-studied nearby systems with well-known periods and orbital elements; the stars with the poorest orbital constraints are in the process of being updated with archival RV and astrometry measurements (Section \ref{sec:targets}, Table \ref{tab:binaries}). The measured binary separations will include contributions from differential LEO orbital-aberration, differential annual-aberration, and differential annual parallax (the two stars are at different distances, easily resolved at the $\mu$as-level). The former two are known from celestial mechanics to better than the mission measurement floor ($\sim$0.1 $\mu$as). For example, for $\alpha$ Cen, SHERA in LEO produces signals of order $\sim$200 $\mu$as with a $\sim$98-minute periodic modulation of precisely known amplitude and phase. Recovery of these signatures at the expected level will provide an end-to-end verification of the astrometric measurement chain and demonstrate that the SHERA pipeline can correctly reproduce known celestial motions when searching for small planetary perturbations. Similarly, the binary differential parallaxes will have well known phases and amplitudes (in the 5-50 $\mu$as range) across the targets. The discovery space most susceptible to degeneracy between uncertainty in the binary orbit and a planetary signal is for the longest period accessible planets ($\sim$6 years, for which SHERA will observe $\sim$half an orbit) around the binary for which we have the poorest knowledge, HR 2667/2668. In this edge case, long-term trends in the data are less obviously attributable to a putative planet.

\section{Precursor Science for the Habitable Worlds Observatory}
\label{sec:hwo}

The 2020 Decadal Survey \cite{Astro2020} highlighted ``Pathways to Habitable Worlds" and ``Worlds and Suns in Context" as two of the primary scientific themes of the coming decade, and outlined a program to identify and characterize Earth-like planets. NASA's response to the Decadal recommendation has been to begin studying the \HWO\ (HWO) mission concept \cite{Feinberg2024}. There has been significant community interest in supporting the formulation of the observatory, both in advancing its architecture-driving science cases and requisite technologies, and there are near-future opportunities to conduct both {\it precursor science} investigations to inform the design of the HWO mission, and longer-term {\it preparatory science} investigations to inform the survey design and enhance the scientific return of HWO. By deeply exploring nearby Sun-like stars, SHERA would pave the way for HWO observations and characterization of nearby planetary systems. The primary SHERA target list (Section \ref{sec:targets}) includes 13 Tier 1 HWO targets and one Tier 2 target \cite{Mamajek2024,Harada2024}. Earth-like planets in any of these systems would be ideally suited for exobiology characterization, due to their similarity to and proximity to the Sun. For example, HWO would receive more photons from an Earth-like planet around $\alpha$ Centauri than from {\it any other possible Earth-like planet}. Although the SHERA data alone will typically not uniquely identify which of the binary stars is the host of the planet, the knowledge that a given system contains an Earth-like planet (or planets) will still be incredibly valuable to HWO planning.

\begin{figure}
\begin{center}
\begin{tabular}{c}
\includegraphics[height=7cm]{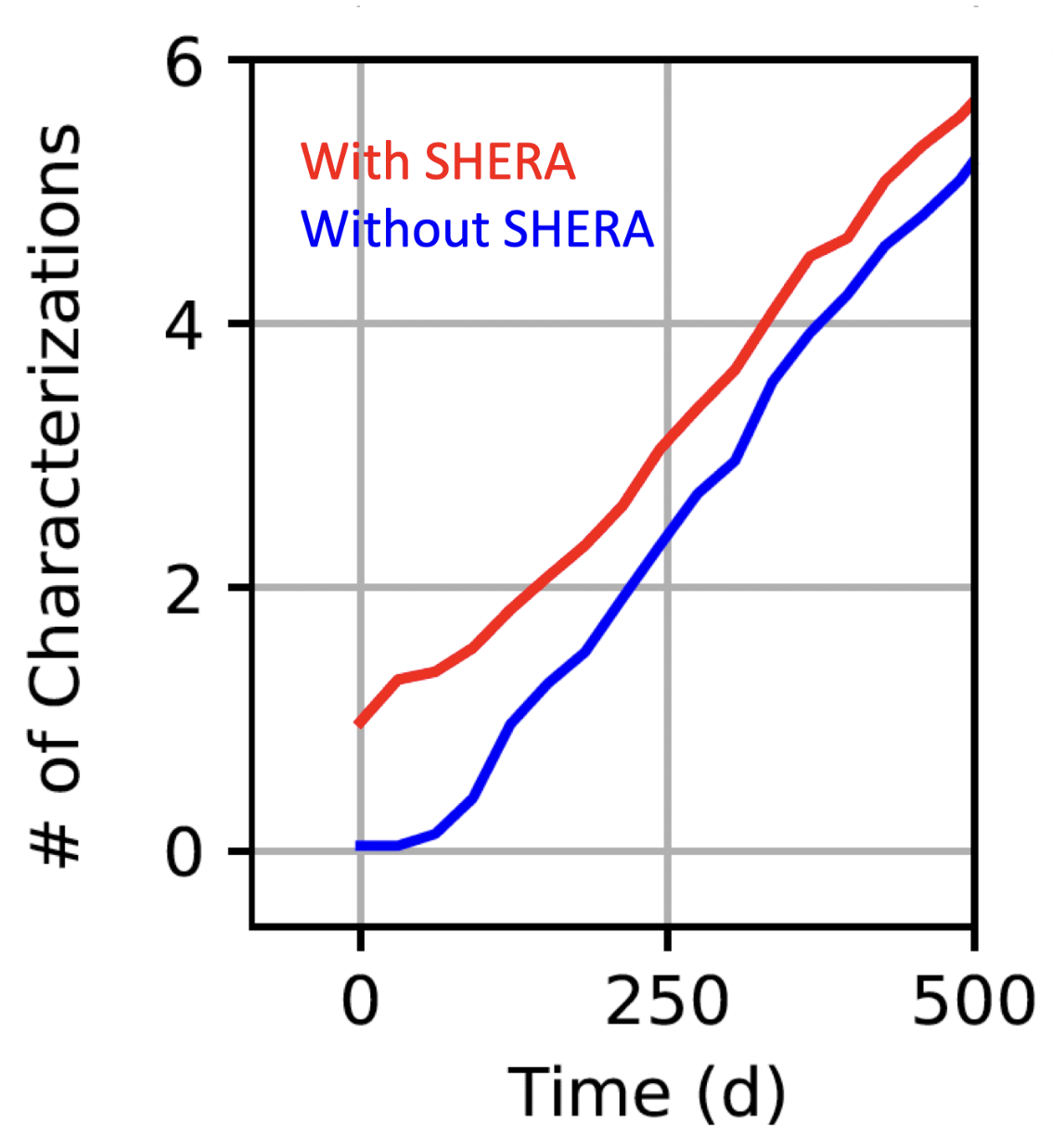}
\end{tabular}
\end{center}
\caption 
{For a simulated example where SHERA finds three planets around high-priority HWO targets, the amount of HWO observing time needed to characterize a given number of planets decreases by up to 40\%.
\label{fig:hwoyield}}
\end{figure}

\cite{Morgan2021} show the benefits of prior knowledge for an exoplanet imaging mission like HWO. Knowing {\it a priori} the existence and orbital parameters for planets discovered by SHERA, HWO would be able to optimize its survey strategy, avoiding a blind search phase for these targets. Using EXOSIMS \cite{Savransky2017} to simulate an example where SHERA finds three planets around high-priority HWO targets, the time needed to characterize a given number of planets is decreased by up to 40\% (Figure~\ref{fig:hwoyield}). In addition, the mass constraints provided by SHERA in combination with the orbital information provided by HWO monitoring would significantly enhance our ability to perform habitability and exobiology studies with the HWO spectra \cite{Damiano2022}. If no planets are found in a deep search of a given SHERA/HWO target, it could be excluded or de-weighted in future survey planning.

Finally, SHERA’s measurement of the occurrence rate of planets in multi-star systems would inform prioritization of those systems in the final HWO target list. Currently, 54 of the 164 Tier 1 HWO targets (33\%) identified by \cite{Mamajek2024} as potentially having the most accessible Earth-like planets for a direct imaging survey are in binary or higher-order multi-star systems. Managing starlight suppression for imaging of multiple bright stars in the field-of-view \cite{Belikov2025} is a technical challenge, the importance of which would be addressed by the survey undertaken by SHERA. Given the importance and urgency of precursor and preparatory science needed for HWO, including characterization of the putative target lists 
\cite{Mamajek2024,Harada2024,Tuchow2024}, especially for assessing habitability and interpreting biosignatures \cite{Meadows2018}, it is crucial that this work start as soon as possible. Besides HWO, SHERA would provide important targets for other mission concepts under consideration, such as the Hybrid Observatory for Earth-like Exoplanets (HOEE)\cite{Soliman2025}, and the Large Interferometer For Exoplanets (LIFE)\cite{Quanz2022}, as well as the next generation of extremely large ground-based optical telescopes.

\section{Conclusions}
\label{sec:conclusions}

Given that nearly half of Sun-like stars are in multi-star systems, understanding their capacity for hosting rocky, habitable zone planets is an essential, and thus far unaddressed, ingredient in understanding the total number of habitable planets in our Galaxy. Nearby Sun-like stars in intermediate-separation (20--200 au) multi-star systems represent a unique opportunity for searching for these planets, by providing built-in metrology scales against which to measure relative astrometry. SHERA is a Small Explorer-class mission concept for a three-year mission, flying a small optical spacecraft in low-Earth orbit, to perform sub-microarcsecond relative astrometry on an initial set of 14 nearby Sun-like stars. With the planned observations, SHERA will be sensitive to rocky planets in the habitable zones of eight stars, i.e. SHERA will be capable of robustly detecting an Earth-sized planet in an Earth-like orbit around a Sun-like star for the first time, one of the most significant current goals in the field. The full survey will be sensitive to a wide range of planets across the period range 80--640 days for all 14 stars, and will be able to statistically constrain how the occurrence rates of these planets differ from those in single-star systems. For any planets that SHERA finds that are amenable to radial velocity follow-up, we can obtain three-dimensional reconstructions of the angular momentum vectors in the system, an essential element in understanding the formation and evolution history of such systems.

\subsection*{Disclosures}
The authors declare that there are no financial interests, commercial affiliations, or other potential conflicts of interest that could have influenced the objectivity of this research or the writing of this paper.

\subsection* {Code, Data, and Materials Availability} 
The work uses a number of publicly available codes, including \texttt{RVSearch} \cite{Rosenthal2021}, \texttt{orbitize!} \cite{Blunt2020}, and \texttt{EXOSIMS} \cite{Savransky2017}. Our modifications to \texttt{orbitize!} will be released in the accompanying paper (Roberson et al., in prep). New radial velocity measurements used to update the binary orbits will be released in the accompanying paper (Clark et al., in prep); the orbits otherwise depend on values in the published literature where referenced.

\subsection* {Acknowledgments}
This research was carried out at the Jet Propulsion Laboratory, California Institute of Technology, under a contract with the National Aeronautics and Space Administration (80NM0018D0004). This research has made use of the NASA Exoplanet Archive, which is operated by the California Institute of Technology, under contract with the National Aeronautics and Space Administration under the Exoplanet Exploration Program.


\bibliography{report}   
\bibliographystyle{spiejour}   

\listoffigures
\listoftables

\end{spacing}
\end{document}